\documentclass[12pt]{article}
\usepackage{amssymb}
\usepackage{amsfonts}
\usepackage{amsmath}
\usepackage{mathtools}  
\usepackage{bm}
\usepackage{latexsym}
\usepackage{epsfig}
\usepackage{bbm}
\usepackage{enumerate}
\usepackage{float}
\usepackage{color}
\usepackage{gensymb}
\usepackage{siunitx}
\usepackage{color}
\usepackage{hyperref}
\usepackage{array,multirow}
\usepackage[round,authoryear]{natbib}
\usepackage[T1]{fontenc}
\usepackage[utf8]{inputenc}
\usepackage{authblk}
\usepackage{appendix}
\usepackage{blindtext}
\usepackage{lineno}
\usepackage{nth}
\usepackage{bbm}
\usepackage{setspace}
\newcommand*\bigcdot{\mathpalette\bigcdot@{.5}}
\providecommand{\keywords}[1]
{
  \small	
  \textbf{\textit{Keywords---}} #1
}

\usepackage[
        a4paper,
        left=2.5cm,
        right=2.5cm,
        top=3cm,
        bottom=3cm]{geometry}

\begin{document}

\title{Estimating Concurrent Climate Extremes: A Conditional Approach}

\author{Whitney K. Huang\footnote{Clemson University. E-mail: \href{wkhuang@clemson.edu}{\nolinkurl{wkhuang@clemson.edu}}}, Adam H. Monahan\footnote{University of Victoria. E-mail: \href{monahana@uvic.ca}{\nolinkurl{monahana@uvic.ca}}}, Francis W. Zwiers\footnote{University of Victoria. E-mail: \href{fwzwiers@uvic.ca}{\nolinkurl{fwzwiers@uvic.ca}}}}

\date{\today}

\maketitle

\begin{abstract} 
Simultaneous concurrence of extreme values across multiple climate variables can result in large societal and environmental impacts. Therefore, there is growing interest in understanding these concurrent extremes.
In many applications, not only the frequency but also the magnitude of concurrent extremes are of interest. One way to approach this problem is to study the distribution of one climate variable given that another is extreme. In this work we develop a statistical framework for estimating bivariate concurrent extremes via a conditional approach, where univariate extreme value modeling is combined with dependence modeling of the conditional tail distribution using techniques from quantile regression and extreme value analysis to quantify concurrent extremes. We focus on the distribution of daily wind speed conditioned on daily precipitation taking its seasonal maximum. The Canadian Regional Climate Model large ensemble is used to assess the performance of the proposed framework both via a simulation study with specified dependence structure and via an analysis of the climate model-simulated dependence structure.
\end{abstract}
\keywords{Concurrent wind and precipitation extremes; quantile regression; conditional extreme value model; large climate ensembles}

\doublespacing
\section{Introduction} \label{sec1}
Concurrent extreme events are the simultaneous occurrence of extreme values for multiple variables. Such events belong to the class of so-called compound extremes \citep{leonard2014,zscheischler2018,hao2018,tilloy2019}. Environmental examples include coastal flooding due to extreme coastal water level and heavy precipitation \citep{van2015,wahl2015,moftakhari2017,ridder2018,bevacqua2019}, compound drought and extreme heat \citep{zscheischler2017}, and concurrent wind and precipitation extremes \citep{martius2016,waliser2017}. Recent examples of events characterized by concurrent extremes include the Russian heatwave (2010; co-occurrence of drought and heat) and Hurricane Sandy (2012; simultaneous heavy rainfall and storm surge). Compound events produce some of the largest impacts on both human society and environmental systems and therefore it is crucial to properly assess the risk of such events \citep{zscheischler2018}.

Estimating compound/concurrent events has become a very active research topic in recent years. Most of the existing work in the literature focuses on estimating the occurrence \textit{probability} of a compound event, defined as two (or more) climate variables exceeding some high percentile threshold \citep{van2015,waliser2017, zscheischler2017,ridder2018,bevacqua2019}. For example, \cite{toreti2019} used the marked inhomogenous J-function statistic \citep{van1999} to summarize the frequency of concurrent heat stress and drought events. \cite{zhou2018} applied copula methods \citep{joe1997,nelsen2007} on the full temperature and precipitation data to estimate the probabilities of concurrent extreme temperature and precipitation over China. The more recent study of \cite{poschlod2020} used a large initial-condition ensemble combined with empirical quantiles to study the occurrence of heavy rainfall on saturated soil during the summer months and concurrent heavy rainfall and snow-melt. 

However, there are also some applications that require the estimation of the \textit{``magnitude''} of concurrent extreme events. For example, in the National Building Code of Canada \citep{national2015}, a critical engineering design value is the Driving Rain Wind Pressure \citep[DRWP,][]{jeong2020}, which is defined as the 5-year return level for wind pressure when it is raining heavily (i.e., more than 1.8 mm/hour). This is an important number for designing the building envelope to ensure that it protects against the possibility of strong winds driving rain water behind the exterior cladding of the building. It is important to note that  one-to-one mapping between the probability of an event and its magnitude given the cumulative distribution function only holds for the unvariate setting. Under the bivariate setting, a given exceedance probability will correspond to a collection of values for both variables (i.e., a bivariate contour), that is, $\{x_{1} \text{ and } x_{2}: F(x_{1}, x_{2}) = 1-p\}$. See, for example, Fig.\ 3 in \cite{cooley2019non}. 

In this work we focus on developing a statistical framework for estimating the magnitude of concurrent extremes with an application to concurrent wind and precipitation extremes. Specifically, we consider methodologies to model the conditional dependence of daily wind speed quantiles on simultaneous block-maxima of daily precipitation. \cite{martius2016} applied a logistic regression \citep{Agresti2013} to ERA-Interim reanalysis products \citep{dee2011} to quantify the odds of having an extreme precipitation  event given that an extreme wind event has occurred, where both the extreme precipitation and extreme wind events were defined as the values exceeding their local seasonal $98\text{th}$ percentiles. They found that the odds of such events have increased in coastal regions and in areas with frequent tropical cyclones. The present study broadens the scope of \cite{martius2016}, which characterize the occurrence probability of  ``moderate'' wind and precipitation extremes \citep{zhang2011}, using methods based on extreme value theory \cite[EVT,][]{fisher1928, gnedenko1943, gumbel1958, davison1990,coles2001, katz2002, heffernan2004} and quantile regression \citep{koenker1978} to model the wind and precipitation tail distributions.

Over the past couple of decades, EVT-based statistical methods have been widely used in climate studies to estimate extremes (i.e., the upper or lower tail distribution) of a single climate variable. In such analyses, one fits a \textit{generalized extreme value} (GEV) distribution  or \textit{generalized Pareto} (GP) distribution to block maxima or threshold exceedances respectively to infer the so-called r-year return level \citep[e.g.,][]{zwiers1998,palutikof1999,kharin2005,jagger2006,cooley2007,cooley2010,kharin2013,westra2013,huang2016,wang2016,risser2017,huang2019a,russell2019,zhu2019,russell2020}. One main advantage of these EVT-based methods is that, in addition to estimating the exceedance probability of a given ``large'' value, one can quantitatively characterize the entire tail distribution (i.e., estimate the exceedance probability for \textit{any} given ``large'' value and high quantiles) given that the underlying theory can provide a reasonable approximate to the extreme values of interest (as discussed in Sec.~\ref{marginal}). However, it is not sufficient to quantify concurrent extremes by conducting pairs of univariate analyses (i.e., only analyzing each variable separately) as doing so could lead to under- or overestimation of risk if the variables of interest are respectively positively or negatively related to each other.            

\begin{figure}[H]
    \centering
    \includegraphics[width=6.5in]{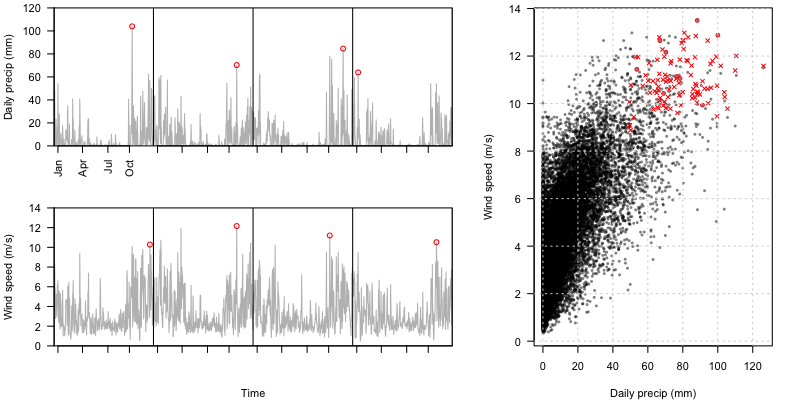}
    \caption{\textbf{Left}: Daily values and annual maxima for precipitation (\textbf{upper}) and wind speed (\textbf{lower}) of the first four years at a Pacific Northwest coastal region grid cell from the output of a climate model simulation (see Sec.~\ref{sec2} for more details). \textbf{Right}: Simultaneous daily data for precipitation and wind speed (black dots) for 100 simulation years. Component-wise annual maxima for these simulation years are shown in red (denoted by `{\Large$\boldsymbol{\cdot}$}' if annual maximum daily precipitation and wind speed happened simultaneously, and by `{\small$\times$}' otherwise). Only 9 (out of 100) component-wise annual maxima occur simultaneously at this grid cell.}
    \label{fig:Fig1}
\end{figure}

There are relatively few climate studies considering the distributions of extremes in a multivariate setting despite a large body of work in the statistical community having been dedicated to modeling multivariate and spatial extremes \citep{tawn1988, tawn1990, smith1990, coles1991,ledford1996, ledford1997,cooley2006,naveau2009,davison2012,wadsworth2012,huser2014,wadsworth2018,huang2019b,wadsworth2019,huser2019,beranger2019,bopp2020}. The existing methods for modeling multivariate (including spatial) extreme distributions mostly focus on ``component--wise extremes'', in which extreme values for each component (e.g., climate variable) are first extracted separately and then combined to create a new extremal data vector (see, for example, Chapter 8.2 of \cite{coles2001} and \cite{davison2012} for a case study that models annual maximum precipitation at several weather stations in Switzerland). This process is illustrated in Fig.~\ref{fig:Fig1}. A drawback of this data selection process (and the resulting statistical analyses) is the neglect of the information regarding the timing of the extremes of the individual variables: the extreme events for the different variables do not necessarily occur simultaneously, and the extremal data vectors do not generally correspond to observed states of the system. In contrast, simultaneity within a time window is a key aspect of compound extreme events and the impacts that they produce.  

\begin{figure}[H]
    \centering
    \includegraphics[width=6.5in]{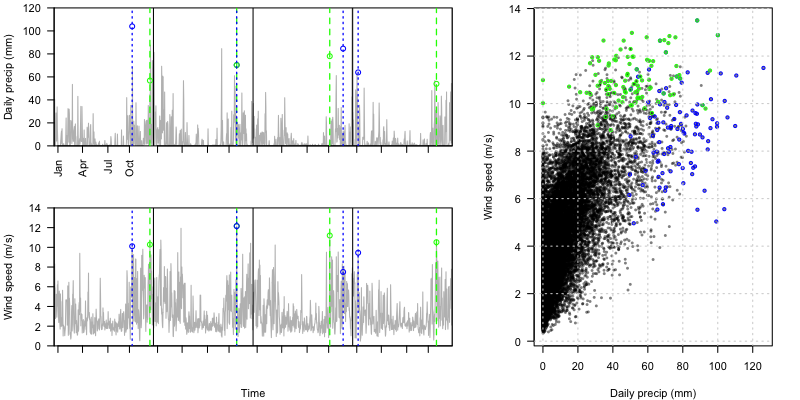}
    \caption{\textbf{Left}: As in Fig.~\ref{fig:Fig1} but now selecting simultaneous data pairs conditioning on precipitation being extreme (blue) or daily mean wind speed being extreme (green). \textbf{Right}: As in Fig.~\ref{fig:Fig1} but including concomitants of maxima (precipitation conditioned on wind speed maxima in green, and wind speed conditioned on precipitation maxima in blue).}
    \label{fig:Fig2}
\end{figure}

One way to overcome the temporal mismatch of extreme values across different climate variables is to frame the problem in terms of conditional concurrent extremes. Instead of using the component--wise extremes, an approach to studying concurrent extremes first obtains the extreme values (e.g., block maxima or threshold exceedances) from a ``conditioning'' variable and then obtains the values of the other variables at the times at which the extreme values occurred, which we denote as the concomitants (see Fig.~\ref{fig:Fig2} for an example). In this way, the selection of points respects the dependence structure of the two variables.
The component-wise extremes arise as a special case if extreme values across different variable occur simultaneously. \cite{dombry2018} addressed the estimation of probabilities of such events. The present study models the {\it distribution} of the conditional extremes.               

\begin{figure}[H]
    \centering
    \includegraphics[width=3in]{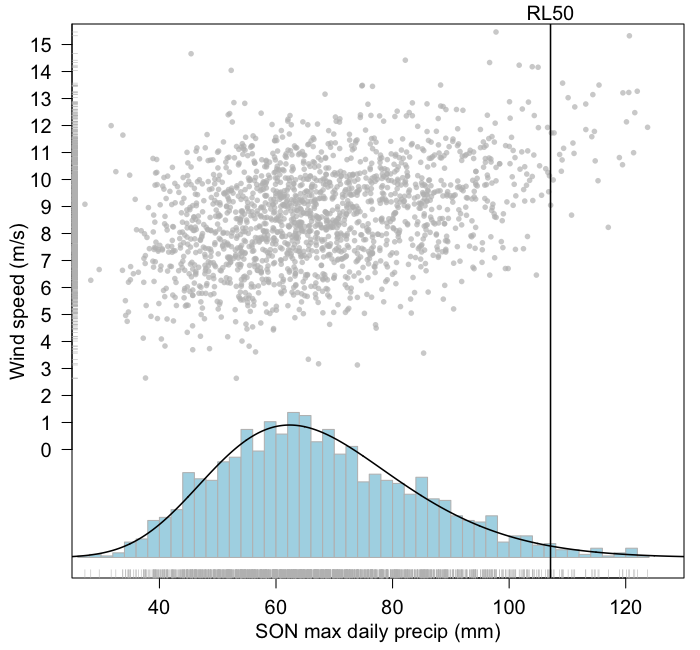}
    \caption{The histogram of annual values of the SON maximum of daily precipitation values at a Pacific Northwest coastal region grid cell from 1950 to 1999 (for all the ensemble members in the dataset, see Sec.~\ref{sec2} for more details) and the associated GEV probability density function estimate (back curve, the vertical line is the estimated 50-year return level). Gray points show the scatter of the SON maximum precipitation and their wind speed concomitants.}
    \label{fig:Fig3}
\end{figure}

This definition of concurrent extremes in terms of their \textit{concomitants} \citep{barnett1976,nagaraja1994} naturally leads to a conditional modeling approach. Specifically, we decompose the estimation of a multivariate (bivariate in this work) distribution given the conditioning variable being extreme into to i) the estimation of the tail distribution of the conditioning variable, and ii) the estimation of the conditional distribution \textit{given} the conditioning variable being extreme. Step i) is the classical problem in extreme value analysis allowing use of techniques from univariate extreme value theory (see Sec.\ 3.1 for more details). Step ii), however, is more challenging because, unlike the step i), there is no general theory to provide a candidate distribution for modeling the conditional distributions. Fig.~\ref{fig:Fig3} gives an illustration of the step i) where we fit a GEV distribution to climate model simulated annual ``autumn'' (the months of September, October, and November, SON hereafter) maxima of daily precipitation for a coastal area of the North American Pacific Northwest region, which will be the conditioning variable in this case. What is left is step ii): estimating how the wind speed distribution changes with the SON seasonal maxima of daily precipitation values. Such dependence is evident in the scatterplot presented in Fig.~\ref{fig:Fig3}.    
Here we explore two statistical methods to this end: 1) a neural network-based non-parametric quantile regression method \cite[monotone composite quantile regression neural network, MCQRNN hereafter;][]{cannon2018} in which the concomitants of extremes are regressed on the conditioning variable to flexibly estimate a range of conditional quantiles (see Sec. 3.2.1 for more details); 2) the semi-parametric conditional extreme value (CEV) approach proposed by \cite{heffernan2004} (see Sec.\ 3.2.2 for more details). 

A limitation of both methods is that conditioning on one of the two variables breaks the symmetry of the concurrent extreme analysis: the results of the analysis may be different if the roles of the conditioning and conditioned variables are exchanged. While in some applications the choice of conditioning and conditioned variables may be clear (e.g., extremes of temperature given that a drought is occurring, or the distribution of wind speed given extreme precipitation in order to estimate extreme DRWP for building design) in other applications this may not be the case. The main reason for taking the conditional approach here is so that we can combine efficient parametric estimation of the conditioning marginal variable with a more flexible representation of the dependence structure than would otherwise be difficult with a full parametric model. 

To develop our conditional extreme analysis with a focus on concurrent wind and precipitation extremes, we use a large ensemble of simulations from the Canadian Centre for Climate Modelling and Analysis Canadian Regional Climate Model 4 (CanRCM4 hereafter, see Sec.\ref{sec2} for more details). Large initial condition climate change ensembles, which are produced with a single climate model under a particular radiative forcing scenario by using differently perturbed initial
conditions for each member of the ensemble to create a set of approximately independent realizations of internal variability \citep{deser2012, kay2015, sriver2015}, offer several advantages over single climate model simulations. One clear benefit of having a large ensemble is that one can obtain more precise estimates of nonstationary extreme statistics (by treating all simulations as approximately statistically independent to each other due to
the sensitivity of these models to atmospheric initial conditions) than with just a single model run \citep{haugen2018,haugen2019}, which is crucial given that the amount of ``extreme'' data is limited by the very definition. This aspect of the large ensemble is not the focus of the present study, but will be addressed further in the discussion. The second benefit is that a large ensemble provides a means to assess the performance of statistical methods \citep{stein2020}. We will use this second aspect to empirically assess the estimation performance of the quantile regression and the conditional extreme methods in the analysis of concurrent wind and precipitation extremes.

The remainder of this paper is structured as follows: in Sec.~\ref{sec2}, we describe the output from the large climate model ensemble simulation used in this study; in Sec.~\ref{sec3}, we provide background for the univariate extreme value analysis (Sec.~\ref{marginal}), quantile regression (Sec.~\ref{qr}), and conditional extreme value models (Sec.~\ref{cevm}) we employ. We also describe how we can combine these tools to estimate the distribution of concurrent extremes. A simulation study is presented in Sec.~\ref{sec4} and Sec.~\ref{sec5} presents an analysis of wind and precipitation concurrent extremes. Section~\ref{sec5} also shows how the large ensemble can be used to assess the performance of the methods considered when only single climate realizations are available (as in the observational record). We conclude with a discussion of the implications of these results.

\section{CanRCM4 Large Ensemble} \label{sec2}

The large ensemble used in this study is produced by the Canadian Regional Climate Model (RCM) version 4, CanRCM4 \citep{scinocca2016}. Each member of the ensemble of RCM simulations was driven by a corresponding member of a large ensemble of simulations of its parent global climate model (GCM), the second generation Canadian Earth System Model (CanESM2), for which the Canadian fourth generation Atmospheric Model (CanAM4) forms the atmospheric component. The RCM was run at $0.44^{\circ} \times 0.44^{\circ}$ horizontal grid resolution ($\sim$ 50 km) over the North American domain defined by the Coordinated Regional Climate Downscaling Experiment (CORDEX) project (\url{https://www.cordex.org/domains/region1-north-america/}). The resolution of this ensemble is finer than most other available large ensembles. While this relatively high resolution is sufficient to capture processes associated with synoptic-scale variability, it is still too coarse to allow the model to represent mesoscale convective systems (e.g., convective storms). Nonetheless, an evaluation study reported in \cite[][Sec. 4.1]{jeong2020} suggests there is a reasonable agreement between  Canadian station observations and CanRCM4 ensemble averages of precipitation and wind speed for the 1986–2016 period.

\begin{figure}[H]
    \centering
    \includegraphics[width=3in]{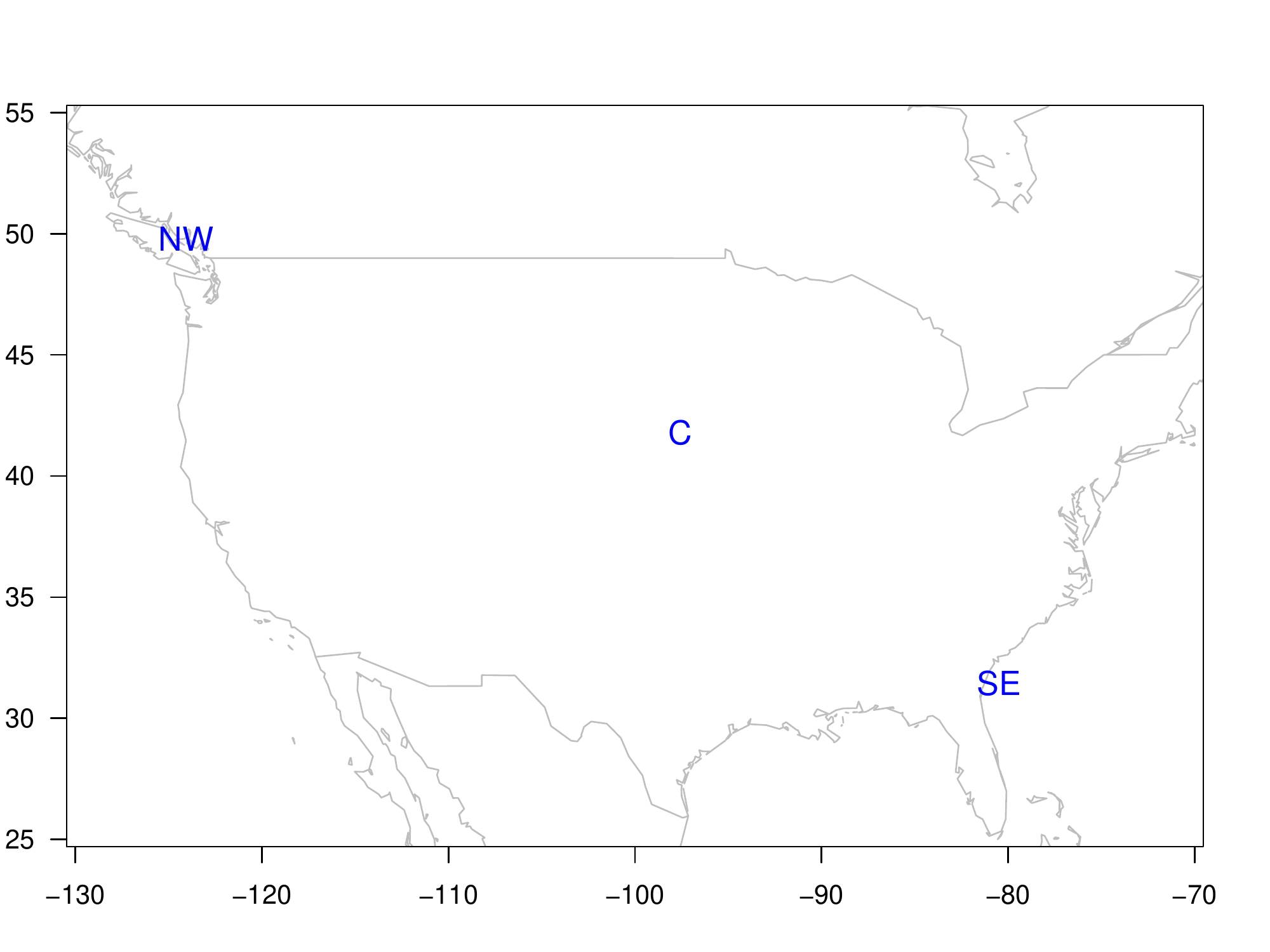}
    \caption{The three grid cells considered in this analysis: Pacific Northwest (NW), Continental interior (C), and Southeast (SE).}
    \label{fig:map}
\end{figure}

The CanRCM4 large ensemble \citep{fyfe2017,kirchmeier2017,kirchmeier2019,li2019a,li2019b} contains 50 members with simulations spanning from 1950-2100 driven by the CanESM2 large ensemble, using historical forcing from the Coupled Model Intercomparison Project Phase 5 (CMIP5) for 1950-2005 and Representative Concentration Pathway (RCP) 8.5 forcing from 2006 to 2100.   
In this work, we use 35 ensemble members providing 3-hourly output averaged to provide daily precipitation and daily mean wind speed. To demonstrate the proposed framework we analyze simulation results at a Pacific Northwest coastal region grid cell (NW), a continental interior grid cell (C), and a Southeast Atlantic coastal ocean grid cell (SE; Fig.~\ref{fig:map}). We consider CanRCM4 output for four seasons: December, January, February (DJF); March, April, May (MAM); June, July, August (JJA); and SON. Because discernible non-stationarity is evident especially for the daily precipitation distribution under the RCP 8.5 forcing scenario (see Fig.~\ref{fig:nonstatTs} in Appendix A.2), we only consider model output from 1950-1999 over which the responses to the anthropogenic forcing in precipitation is relatively small. Non-stationarity in simulated 10-meter wind speed is less evident, partly because the land surface properties (in particular roughness length) were held constant in the CanRCM4 integrations. Focusing on the first 50 years of the simulation mitigates the complications due to nonstationarity in the marginal distributions and potentially in the dependence structure as well, which would require the development of a nonstationary version of the our proposed framework.     

\section{Statistical Framework} \label{sec3}
In this section we describe a conditional approach to modeling the distribution of concurrent extremes, which consists first of the estimation of the tail distribution of the conditioning variable (marginal modeling), and second of dependence modeling where the conditional distribution of one variable given the another variable being extreme needs to be estimated. Let $\bm{X} = (X_{1}, X_{2})^{T}$ be the random vector of interest and let $(x_{1,i}, x_{2,i}), i = 1, \cdots, m$ be a realization of $\bm{X}$. We describe the marginal tail estimation (i.e., estimating high quantiles of $X_{1}$, $X_{2}$) in Sec. \ref{marginal} and our strategy for dependence modeling in Sec. \ref{dependence}.

\subsection{Marginal modeling} \label{marginal}

The goal of marginal modeling is to estimate the tail distribution of a given variable, which is the main focus of the classical extreme value analysis \citep{gumbel1958,coles2001}. Here we give a brief review of the two standard extreme value methods: block maxmia (BM) and peaks-over-threshold (POT). 

The BM method involves fitting a GEV distribution to block maxima, where the blocks are produced by dividing the data record (e.g., a time series of certain climate variable) into non-overlapping periods. The theoretical justification of the BM method is based on the extremal types theorem \citep{fisher1928,gnedenko1943}, which states that, under some regularity conditions of the parent distribution $X$, the distribution of the block maxima $M_{n} = \max_{i=1}^{n}X_{i}$, where $X_{i} \stackrel{i.i.d.}{\sim } F$ (i.e., independent and identically distributed variables) converges to a GEV distribution as the block size $n$ increases. The independence assumption can be
relaxed for weakly dependent stationary time series \citep[e.g.][]{leadbetter1983} and \cite{einmahl2015}
extended the theory to non-identically distributed observations
when distributions share a common absolute maximum). The POT method, which is justified by the same set of conditions of the BM method, involves fitting a GP distribution to data above a high threshold, $u$, given that the threshold is sufficiently high \citep{pickands1975}. While the POT method has an advantage over the BM method in that it typically makes use of the available data more efficiently in estimating extreme events, 
it also has disadvantages such as being more affected by the seasonality and the temporal dependence found in most climate records \citep{katz2002}. Another practical issue when implementing the POT method is
the need to choose the threshold $u$; the high quantile (e.g.\ r-year return level) estimates may be sensitive to the chosen threshold \citep{scarrott2012, wadsworth2012a,huang2019a}. Although there have been some recent attempts \citep{huang2019a,tencaliec2019} to directly estimate high quantiles of climate variables with Pareto tails (e.g., daily precipitation amount in some regions) so that the threshold selection can be avoided.

This study will nevertheless focus on marginal extremes defined by the BM method because using the GEV distribution to model block maxima is reasonably robust with respect to the i.i.d. assumption. The underpinning asymptotic theory from which the GEV distribution emerges continues to hold when applied to data from stationary processes that satisfy suitable mixing conditions \citep{leadbetter1983}. In this case the effect of temporal dependence is that the effective block size will become smaller than the nominal block size raising the issue of whether that effective block size is ``large'' enough so that the GEV distribution can still provide a reasonable approximation. The identically distributed assumption can be somewhat relaxed provided there is quasi-stationarity during the season or subseason when the annual maximum typically occurs, which is an assumption that is implicitly made in all applications of the BM method to meteorological data \citep[e.g.,][]{katz2002}. In our implementation, we estimate the GEV paramters $(\mu, \sigma, \xi)$ of the conditioning variable, the annual maximum daily precipitation for a given grid cell, using maximum likelihood estimation by assuming these annual maximum are independent realizations of a GEV random variable.

\subsection{Estimation of the conditional distribution} \label{dependence}

The key step in our conditional approach is to estimate $[X_{j}|X_{i} \text{ large}], \quad i,j = 1,2, i \neq j$, the conditional distribution of $X_{j}$ given that $X_{i} =x_{i}$ is large (i.e.\ $F_{X_{i}}(x_{i}) = 1 - 1/r$ with large $r$). The main challenge for this task results from the fact that, unlike the univariate setting where one can use extreme value theory to fit asymptotically justifiable distributions to extreme data, there is no general result for the form of the desired conditional distribution. We therefore explore two different methods for estimating the conditional distributions, namely the monotone composite quantile regression neural network (MCQRNN Sec.~\ref{qr}) and conditional extreme value models (CEV, Sec.~\ref{cevm}).

\subsubsection{Monotone composite quantile regression neural network} \label{qr}

The basic idea of quantile regression is to extend the scope of classic regression analysis, which models how the \textit{mean} of a response ($\mathbb{E}(Y)$) varies with a set of explanatory variables (i.e., covariates), to modeling how a \textit{quantile} of a response $Q_{Y}(\tau) = F^{-1}_{Y}(\tau) = \inf\{y: F(y) \ge \tau\}, \tau \in [0, 1]$ changes with covariates \citep{koenker1978}. Since the quantile function $\{Q_{Y}(\tau), \tau \in [0, 1]\}$ fully determines the distribution $F_{Y}$, one could estimate a range of quantile functions ($Q_{Y}(\tau_{k}|X=x)$, $\tau_{k} \in [0, 1], k = 1, \cdots, K$, $x \in \mathbb{R}$) to approximate the underlying (conditional) distribution. In the context of concurrent extremes, we are interested in estimating conditional quantiles $Q_{X_{j}}(\tau|X_{i}=x_{i}),\, \tau \in [0, 1] $ conditioning on $X_{i}$ being ``large''. We will explore non-parametric quantile regression over $K$ different quantile values: that is, we will model $\{Q_{X_{j}}(\tau_{k}|X_{i}=x_{i}) = g_{\tau_{k}}(x_{i})\}_{k=1}^{K}$ where we do not impose strong structure assumptions on $\{g_{\tau_{k}}(\cdot)\}_{k=1}^{K}$ other than some smoothness conditions. One commonly encountered issue with the separate estimation of several quantile functions is that quantile curves may cross (e.g., $g_{\tau_{i}}(x) > g_{\tau_{j}}(x)$ for some $x \in \mathbb{R}$ when $0 \le \tau_{i} < \tau_{j} \le 1$) \citep{he1997,bondell2010,mckinnon2020}. Here we explore how MCQRNN \citep{cannon2018}
 avoids this quantile crossing issue. Specifically, \cite{cannon2018}
  models several quantile curves simultaneously using a feedforward neural network while treating
the quantile level $\{\tau_{k}\}_{k=1}^{K}$ as an additional covariate and imposing partial monotonicity constraints \citep{zhang1999feedforward} on its
associated weight parameters to enforce non-crossing. Other non-parametric models of conditional quantiles were considered. In particular, a preliminary analysis comparing
neural network quantile regression \citep[QRNN,][]{cannon2011} with quantile smoothing splines \citep{koenker1994} was made. The results of this analysis indicated that
the bias and variance performance of quantile smoothing splines was no better than -- and in some cases worse than --
QRNN. Nevertheless, MCQRNN represents only one non-parametric approach to modeling conditional quantiles.
A more thorough comparison with other modeling approaches \citep[e.g.,][]{mckinnon2020,xu2021} is an interesting direction of future study.

\subsubsection{Conditional extreme value model} \label{cevm}
We also explore the conditional extreme value (CEV) models first proposed by \cite{heffernan2004} to estimate the conditional distribution of interest. 
In what follows we briefly describe the CEV models in the bivariate context. The first step in the method involves marginal modeling of the full range of the distributions.  \cite{heffernan2004} took a mixture approach where they used an empirical distribution below a chosen threshold $u$ and a GP distribution above that threshold for each variable. They then used the probability integral transformation to transform each marginal $X_{i}$ to variables $Y_{i}, i = 1, 2$ both distributed as a standard \textit{Gumbel} distribution (or alternatively, to the standard \textit{Laplace} distribution in \cite{keef2013}). The main assumption in CEV models is that, given one of the variables is large ($Y_{1}$ without loss of generality), then the conditional distribution (i.e., $[Y_{2}|Y_{1}>u_{1}]$), is independent of the tail distribution of $Y_{1}$ (i.e., $[Y_{1}|Y_{1}>u_{1}]$) after an appropriate standardization. Specifically, the method assumes that, 
\begin{equation}
        \left[\frac{Y_{2} - a(Y_{1})}{b(Y_{1})} \le z| Y_{1} > u_{1}\right] \sim G(z),
        \label{eq1}
\end{equation}
where $a(y)$ and $b(y)$ are standardizing functions of $y$ for $y > u_{1}$. We will discuss how we set the thresholds $u$ and $u_{1}$ in next section. 

\cite{heffernan2004} found that for most standard copula dependence models studied by \cite{joe1997} and \cite{nelsen2007}, the forms of $a(y)$ and $b(y)$ fall into simple classes when using Gumbel marginals, such that the forms can be further simplified when using Laplace margins. \cite{keef2013} assume the functional forms $a(y) = \alpha y,\, \alpha \in [-1, 1]$ (such that $0< \alpha \leq1$ and $-1 \leq \alpha < 0$ correspond respectively to positive and negative association of $Y_{2}$ and large $Y_{1}$) and $b(y) = y^{\beta}, \beta \in (-\infty, 1]$. The CEV models can be considered a class of semi-parametric models where some parametric assumptions are made regarding how the location and scale change with respect to the conditioning variable. The non-parametric aspect comes in 
through the estimation of the ``residual'' distribution $G$. Further details of the CEV model fitting can be found in 
Appendix.~\ref{App:fit}. 

\section{Simulation Study} \label{sec4}

The purposes of the following simulation study are to 1) examine the performance of MCQRNN and CEV methods in a setting in which the conditional dependence is known, and; 2) to describe our model fitting approach which will also be used to describe the CanRCM4 concurrent wind and precipitation extremes considered in the next section.
We simulate realizations from bivariate random variables where their marginal distributions follow the (estimated)
seasonal maximum daily precipitation distribution (i.e., the fitted GEV distribution) and the concurrent wind speed distribution (i.e., the fitted Weibull distribution) for individial seasons for each of the three selected model grid cells (NW, C, SE) from the CanRCM4 ensemble, but with specified conditional dependence. Here we choose SON data at the NW grid cell as there is clear dependence; JJA at the C grid cell as there is weaker but still evident dependence, and DJF data at the SE grid cell where the dependence is weakest. These data are displayed in Fig.~\ref{fig:Sim1} (see Fig.~\ref{fig:histDJF} in Appendix.~\ref{App:hists} for all season and grid cell combinations). 

\begin{figure}[H]
    \centering
    \includegraphics[width=4.5in]{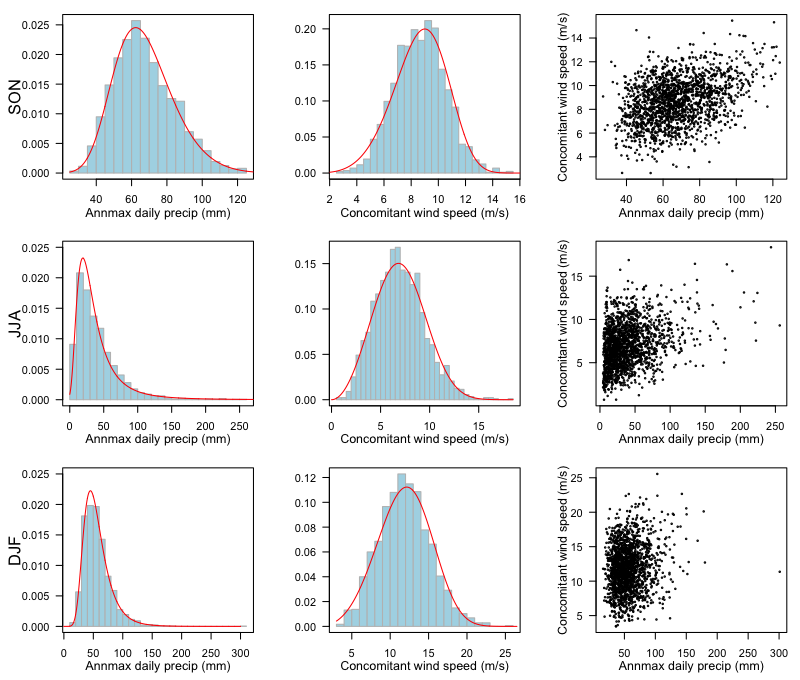}
    \caption{\textbf{Top row}: Histograms of SON maximum daily precipitation (\textbf{Left}), the concomitant daily average wind speed (\textbf{Middle}), and the scatterplot of both together (\textbf{Right}) for a Pacific Northwest (NW) grid cell. \textbf{Middle row}: As in the top row but for JJA maximum daily precipitation for a Continental interior (C) grid cell. \textbf{Bottom row}: As in the top row but for DJF maximum daily precipitation for a Southeastern (SE) grid cell. The red curves are the estimated densities (GEV for annual maximum precipitation and Weibull for concomitant wind speed).}
    \label{fig:Sim1}
\end{figure}

For each grid cell we consider three families of dependence structures: Gaussian, Gumbel, and Clayton copulas \citep{joe1997,nelsen2007}, each with three levels of dependence strength. In each case we generate a Monte Carlo sample of 100 realizations, each with sample size
$n=100$ (corresponding to 100 ``years'' of the simulated data). For the Gaussian copula we set $\rho = (0.1, 0.3, 0.6)$, for the Gumbel copula $C(u_{1}, u_{2}) = \exp\left[-\left\{(-\log u_{1})^{\frac{1}{\alpha}}+(-\log u_{2})^{\frac{1}{\alpha}}\right\}^{\alpha}\right], \alpha \in (0, 1)$ we set $\alpha=(0.5, 0.7, 0.9)$, and for the Clayton copula $C(u_{1}, u_{2}) = \left(u_{1}^{-\delta}+u_{2}^{-\delta}-1\right)^{\frac{-1}{\delta}}, \, \delta>0$ we set $\delta = (0.1, 0.5, 0.9)$. These copula parameters are chosen to represent ``weak'' ($\rho=0.1$, $\alpha=0.9$, and $\delta=0.1$), ``median'' ($\rho=0.3$, $\alpha=0.7$, and $\delta=0.5$), and ``strong'' ($\rho=0.6$, $\alpha=0.5$, and $\delta=0.9$) tail dependence. Also, the three copulas are chosen to examine the effects of different tail dependence structures on the estimation of the conditional upper quantiles (i.e., asymptotic tail independence for Gaussian, positive tail dependence in the upper joint tail for Gumbel, and greater dependence in the lower joint tail than in the upper joint tail for Clayton).

Due to the respectively non-parametric and semi-parametric natures of MCQRNN and CEV, both methods require some ``tuning''. Here we choose the number of hidden layers and hidden nodes to both be equal to 2 to balance the trade-off between bias and variance for the given sample size (more parameters resulting in more flexible models but also more sampling variability). For CEV, we choose $u$, the marginal threshold to switch between the empirical distribution and the GP distribution for the conditioning variable, to be its 0.6 quantile. Note that the conditioning variable, the seasonal maximum daily precipitation, can be reasonably approximated by a GEV so that the threshold does not need to be ``high'' when using a GP distribution to approximate its tail distribution, and the marginal threshold for the concomitant variable to be its 0.75 quantile. In addition, $u_{1}$, the threshold beyond which the location-scale dependence form is assumed, is chosen to be the 0.6 quantile of the conditioning variable (note that this threshold does not have to be the same as the marginal threshold of the conditioning variable). For both methods, we compute the estimated conditional upper quantiles $\bm{\tau} = (0.5, 0.6, 0.7, 0.8, 0.9)$ and compare these with the corresponding true conditional quantiles under the different dependence structures imposed. We perform the estimation for each Monte Carlo sample using a frequentist setting. These same estimation procedure values will be used for the application in the next section. 

\begin{figure}[H]
    \centering
    \includegraphics[width=3in]{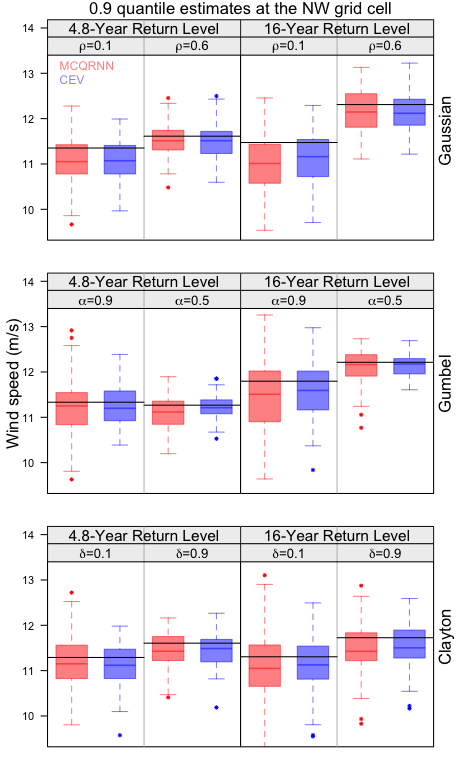}
    \hspace{0.05in}
    \includegraphics[width=3in]{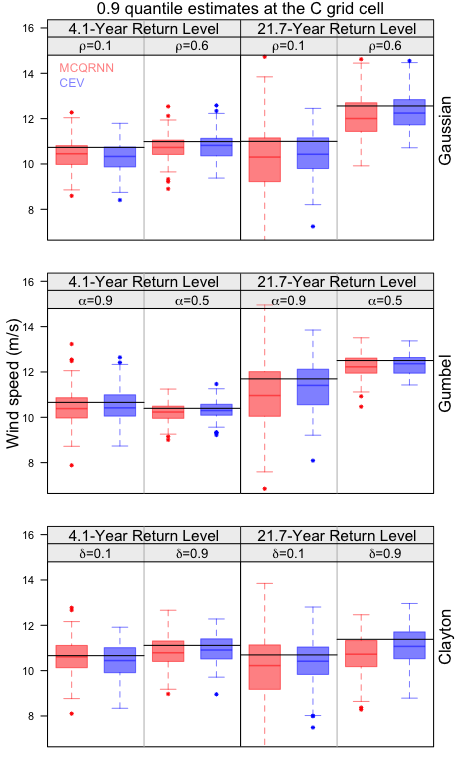}
    \caption{Boxplots of 100-member Monte Carlo 0.9 quantile estimates at the NW \textbf{(Left)} and C \textbf{(Right)} grid cells under three different copulas (\textbf{Top}: Gaussian, \textbf{Middle}: Gumbel, \textbf{Bottom}: Clayton) and two levels of dependence strengths (Weak and Strong, from left to right within each panel) for two different values of the conditioning variable (expressed in terms of return level; low, high from left to right panels). The black horizontal lines are the true 0.9 quantiles.}
    \label{fig:Grid1Cops}
\end{figure}

\begin{figure}[H]
    \centering
    \includegraphics[width=3in]{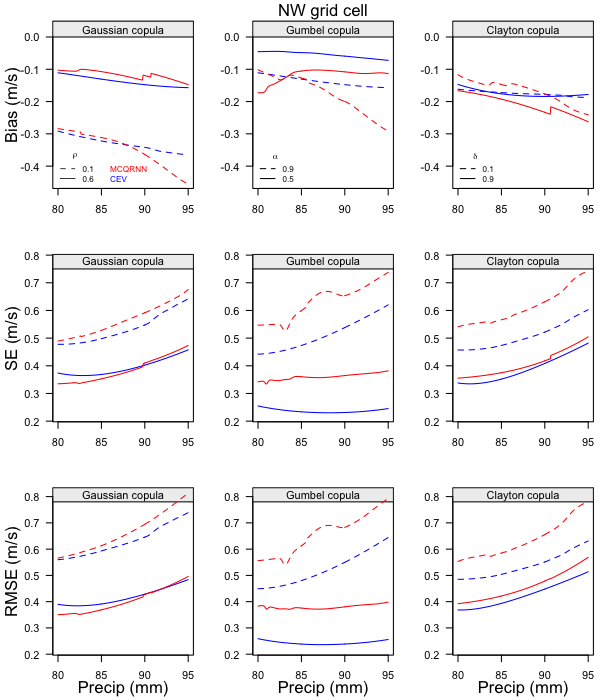}
    \hspace{0.1in}
    \includegraphics[width=3in]{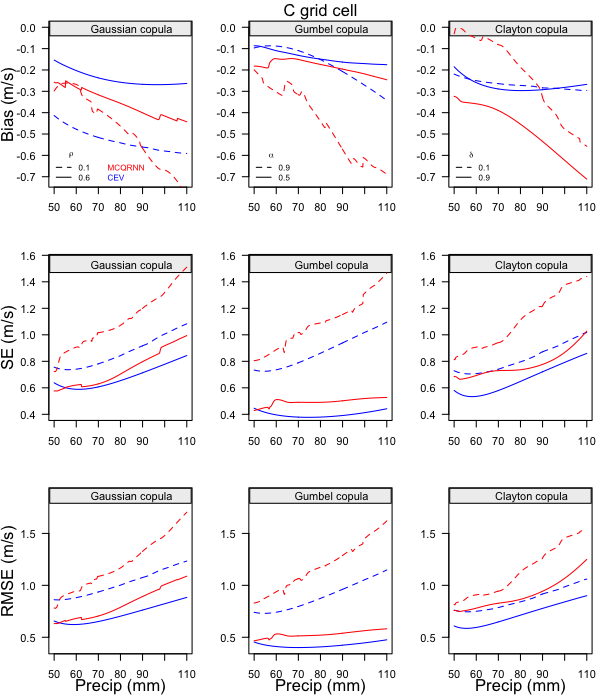}
    
    \caption{Monte Carlo estimates of bias, standard error (SE) and root-mean-square error (RMSE) of conditional quantile values as functions of the value of the conditioning variable (x-axis) for the NW \textbf{(Left)} and C \textbf{(Right)} grid cells under different copula and dependence level combinations.}
    \label{fig:Grid1BiasSe}
\end{figure}

The left panels of Figs.~\ref{fig:Grid1Cops} and \ref{fig:Grid1BiasSe} show the estimates of all the 100 Monte Carlo samples of the highest conditional quantile ($\tau=0.9$) of the NW grid cell during SON for all copulas and dependence levels considered. Both methods have generally similar estimation biases for the Gaussian and Clayton copulas while the CEV outperforms MCQRNN in terms of RMSE under the Gumbel copula dependence structure (Fig.~\ref{fig:Grid1BiasSe}). Note that both methods slightly underestimate the conditional upper quantile curves and that these biases typically increase for more extreme values of the conditioning variable. These negative biases can be largely attributed to the fact that both methods use the \texttt{quantile} command in the \texttt{stats} \texttt{R} base package \citep[see][for more details]{hyndman1996} that leads to underestimation of high quantiles ($\tau=0.9$ in our study). We also found that CEV tends to have smaller bias, presumably due to that fact that MCQRNN estimates high quantiles ``locally'' while CEV estimates quantiles ``globally'' after standardizing with the location and scale parameters (i.e., $\alpha$ and $\beta$) of the dependence structure, and therefore has a larger effective sample to estimate high quantiles (see Appendix.\ E for a simulation study to demonstrate the underestimation using empirical quantiles). In terms of high conditional quantile estimation variability, the CEV estimator tends to have smaller standard error than MCQRNN due to the parametric form of its conditional quantile curves. As with the bias, the standard error of the estimate is typically larger for larger values of the conditioning variable.     

The right panels of Figs.~\ref{fig:Grid1Cops} and \ref{fig:Grid1BiasSe} show the same quantities as the left panels of Figs.~\ref{fig:Grid1Cops} and ~\ref{fig:Grid1BiasSe} for the continental interior grid cell in JJA. The ``summer'' maximum precipitation distribution at this grid cell has a heavier upper tail than that of the autumn NW grid cell (Fig.~\ref{fig:Sim1}). The results show qualitatively similar estimation behaviour: CEV provides more stable estimates and more accurate estimates under the Gumbel dependence structures but again both methods tend to slightly underestimate the target quantile. The accuracy and precision of the MCQRNN estimator degrade more rapidly than for the CEV with increasing values of the conditioning variable, likely due to its non-parametric fitting nature with relatively small number of data points in the far tails. The results for the SE grid cell during DJF are qualitatively similar to those for the C grid cell (not shown).

\begin{figure}[H]
     \centering
     \includegraphics[width=5in]{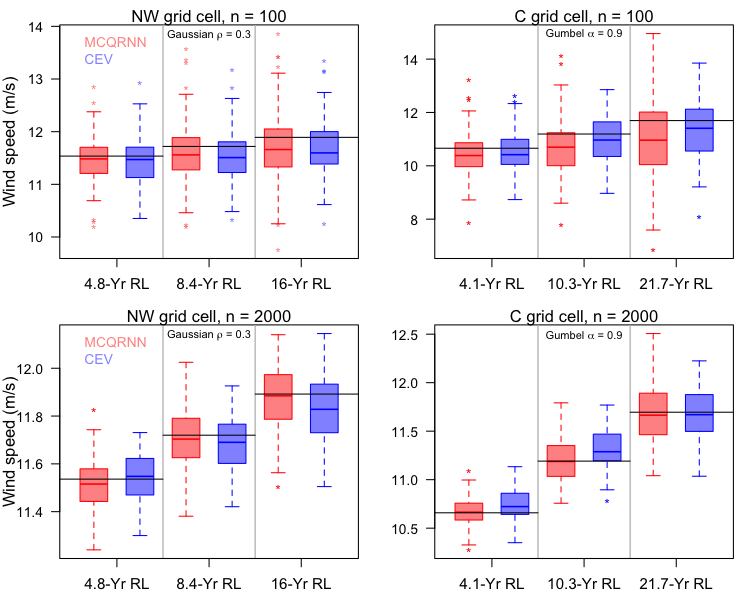}
     \caption{As in Fig.~\ref{fig:Grid1BiasSe} but contrasting results for sample sizes $n=100$ (\textbf{Upper}) and $n=2,000$ (\textbf{Lower}). Results are shown only for NW with Gaussian copula ($\rho=0.3$, \textbf{Left}) and C with Gumbel copula ($\alpha=0.9$, \textbf{Right}). Note that the vertical axes are different across different sample sizes in order to facilitate comparisons between MCQRNN and CEV.}
     \label{fig:sim_n2000}
\end{figure}

We then increased the sample size to 2,000 (corresponding to 2,000 ``years'' of the simulated data) to study how MCQRNN and CEV estimators behave when a much larger data set is available (e.g., as in a large climate model ensemble). Fig.\ref{fig:sim_n2000} contrasts the estimates of the conditional 0.9 quantile for sample sizes 100 and 2,000 for the NW grid cell with Gaussian copula having $\rho=0.3$ and the C grid cell with Gumbel copula having $\alpha=0.9$. The results demonstrate the expected substantial reduction of estimation variability, with reduction factors that are approximately equal to the square root of the factor of the sample size ($\sqrt{20} = 4.47$ in this case). Furthermore, the biases of both methods are substantially reduced (particularly that of the MCQRNN). This finding suggests that with a very large sample such as one of size $n=2,000$, MCQRNN should perform similarly to CEV (with smaller bias due to the non-parametric nature of the method but slightly more variable), assuming that similar behavior holds in the other cases explored. When samples are small (e.g., $n=100$), however, we might expect CEV to perform better since the impact of bias in quantile estimation then dominates in MCQRNN (Appendix E).  

\begin{figure}[H]
    \centering
    \includegraphics[width=5.5in]{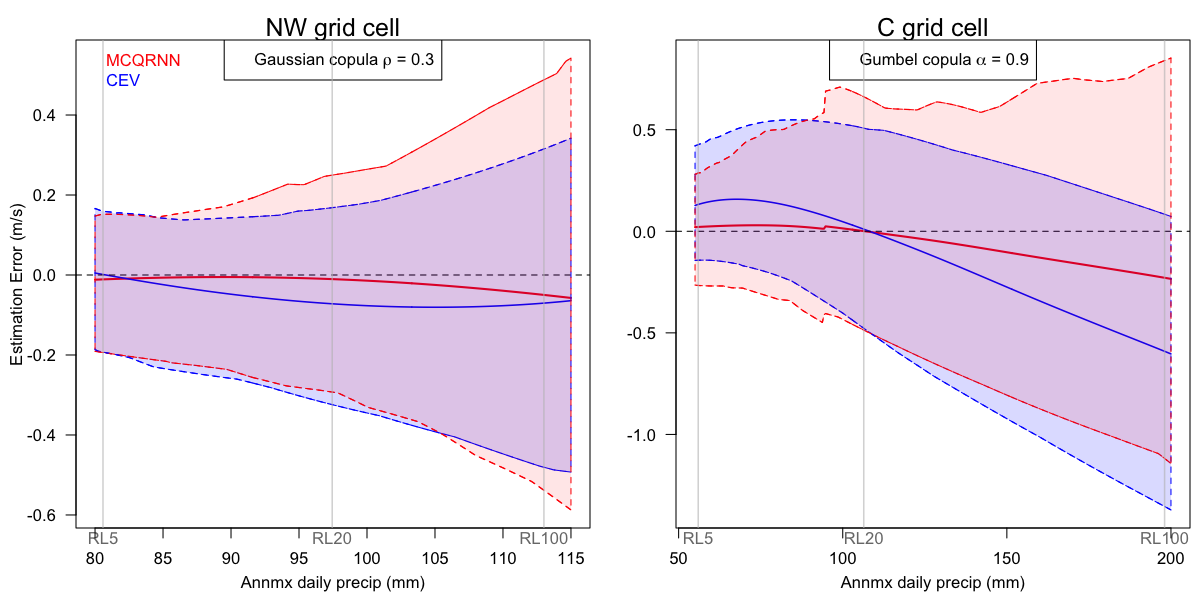}
    \caption{The mean (averaged over 100 Monte Carlo replications) estimation error curve \textbf{(solid)} for the 0.9 conditional quantile and their 2.5/97.5 percentile curves \textbf{(dashed)}. The gray vertical lines indicate the estimated the 5-, 20-, and 100-year return levels of the seasonal maximum precipitation.}
    \label{fig:sim_deeptail}
\end{figure}

A large sample size such as $n=2000$ also allows us to examine the estimation performance for the MCQRNN and CEV methods in the deep tail of the conditioning variable. Fig.~\ref{fig:sim_deeptail} shows the resulting estimates of the conditional 0.9 quantile using MCQRNN and CEV for values of the conditioning variable ranging from the 5 year return level to the relatively extreme 100 year return level. The results indicate while the MCQRNN and CEV methods display similar sampling variability when the value of the conditioning variable is not too extreme, the MCQRNN becomes more variable when conditioning on a large value of the conditioning variable (such as 100-year return level). The increase in sampling variability is particularly pronounced when the conditioning variable is heavy-tailed (e.g., grid cell C) because data are then sparser than for a lighter-tailed distribution (e.g., grid cell NW). These results also indicate that the MCQRNN estimates may be preferable when abundant data are available or not too extreme a value of the conditioning variable is being considered, as its estimates show smaller bias.      

\begin{figure}[H]
    \centering
    \includegraphics[width=5in]{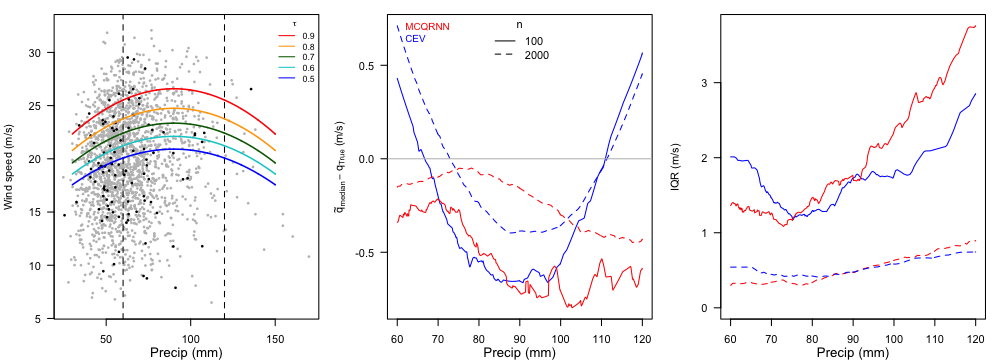}
    \caption{\textbf{Left}: The scatterplot of one realization from the simulated data having non-monotonic extremal dependence with sample size $n=100$ \textbf{(black points)} and $2,000$ \textbf{(gray points)} respectively, with the true conditional upper quantiles shown in colours. The vertical dashed lines indicate the data range for fitting MCQRNN and CEV. \textbf{Middle}: The difference between the Monte Carlo median and the true 0.9 conditional quantile curve for CEV (blue) and MCQRNN (red). The solid lines are for $n=100$ while the dotted lines are for $n=2,0000$.  \textbf{Right}: The interquartile range (IQR) of the 0.9 conditional quantile estimates as functions of the value of the conditioning variable.}
    \label{fig:sim4}
\end{figure}

An aspect of the examples we have considered so far that advantages the CEV is that the true conditional quantile curves for these copulas all have rather ``simple'' structures that can be reasonably well described by parametric CEV model assumptions. Fig.~\ref{fig:sim4} 
therefore shows results obtained in a setting where the conditional quantiles are not monotonically increasing (i.e., the conditional wind speed distribution is Weibull with a scale parameter that is quadratically dependent on the conditioning variable and a fixed shape parameter), and hence cannot be well captured by the CEV model. The resulting bias functions in this case (defined here as the difference between Monte Carlo median and the true value rather than the mean because of large outliers in the MCQRNN estimates) demonstrate that CEV displays a distinct bias function pattern with  overestimation of the conditional upper quantiles in the deep tail, whereas MCQRNN underestimates the conditional upper quantiles. The interquartile range (IQR) functions demonstrate that the CEV estimator is again more stable in the tails of the conditioning variable  compared to that of the MCQRNN, and this difference becomes smaller with larger sample size.            

In summary, both MCQRNN and CEV produce reasonable estimates of conditional high quantiles despite tending to slightly underestimate the true values. When the sample size is around 100 years one may prefer to use CEV as it produces smaller RMSE. On the other hand, if the sample size is ``large'' one may prefer to use MCQRNN as it is less affected by model assumptions and therefore has a smaller bias.
Both MCQRNN and CEV perform less well when there is a non-monotonic dependence  structure, in particular CEV, due the model mis-specification that then occurs.

\section{Wind speed conditioned on annual maximum precipitation} \label{sec5}
In this section we apply our conditional modeling framework to estimate conditional upper quantiles of wind speed \textit{given} precipitation taking its seasonal maximum, using the daily CanRCM4 output at the selected grid cells (NW, C, SE). We first extract the annual maximum  precipitation and the concurrent wind speed values at each grid cell for each season. We then fit GEV distributions to the seasonal annual maximum daily precipitation values for each season at each grid cell. Finally, we estimate the conditional quantiles at $\tau=(0.5, 0.6, 0.7, 0.8, 0.9)$ of the concurrent daily wind speeds using both MCQRNN and CEV.

\begin{figure}[H]
    \centering

    \includegraphics[width=6in]{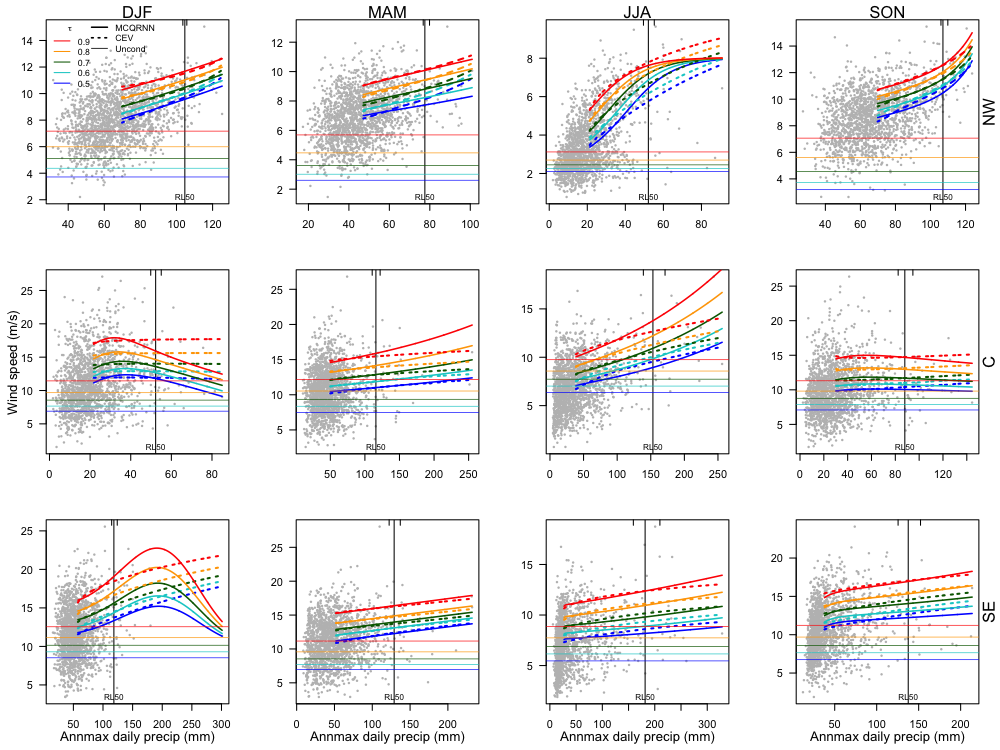}

    \caption{The estimated conditional 0.5, 0.6, 0.7, 0.8, and 0.9 quantiles of wind speed at the NW (top row), C (middle row), and SE (bottom row) grid cells using MCQRNN (thick solid lines) and CEV (dashed lines) for DJF, MAM, JJA, and SON and their corresponding ``unconditional'' upper quantiles (thin solid lines). The vertical lines are estimates of the 50-year return level of daily precipitation (1950-1999, under the stationarity assumption on these marginal distributions) and the vertical ticks on the upper axis around these 50-year return levels indicate the 95 \% confidence intervals based on profile likelihood. The data from which the estimates are obtained are shown in the scatter of gray dots.}
    \label{fig:Fig4}
\end{figure}

Fig.~\ref{fig:Fig4} shows estimates of seasonal conditional upper quantile functions for both MCQRNN and CEV methods. The figure also shows the corresponding estimates of seasonal unconditional quantiles of daily wind speed to assess the influence of conditioning on seasonal maximum daily precipitation.
The results suggest the high quantiles (e.g., 0.9 quantile) of the concomitant wind speed becomes substantially larger than that of the unconditional counterparts, as would be expected when there is positive tail dependence, especially for NW and SE grids, which are both coastal grid points.

Substantial variation is evident across seasons. In general, the wind speeds are larger during SON and DJF for all three locations. Furthermore, at the NW grid cell, SON and DJF tend to have more intense seasonal maximum daily precipitation values relative to MAM and JJA, consistent with the impact of atmospheric rivers, which are well simulated in CanRCM4 \citep{whan2016} and in its driving GCM, CanESM2 \citep{tan2020}. In contrast, more intense seasonal maximum daily precipitation values for the C grid cell are found during JJA and MAM than DJF and SON, reflecting the role of warm season convective systems.

Changes of the dependence structure  across seasons are generally smooth. Both methods produce reasonably close estimates of the conditional upper quantiles of wind speed up to the 50-year return level of seasonal maximum precipitation. Conditional high quantile estimation becomes less robust and less accurate for both methods when the conditioning variable (i.e., seasonal maximum precipitation) becomes more extreme. For example, at the SE grid cell during DJF, a single seasonal maximum precipitation outlier value results in the more flexible MCQRNN deviating substantailly  from the more constrained CEV. Depending on the confidence one has in the veracity of the outlying value, it may be preferred to allow it to have more influence on high quantile estimates by using MCQRNN to make those estimates, or to prefer a weaker influence by using CEV to make those estimates.

\begin{figure}[H]
    \centering
    \includegraphics[width=6in]{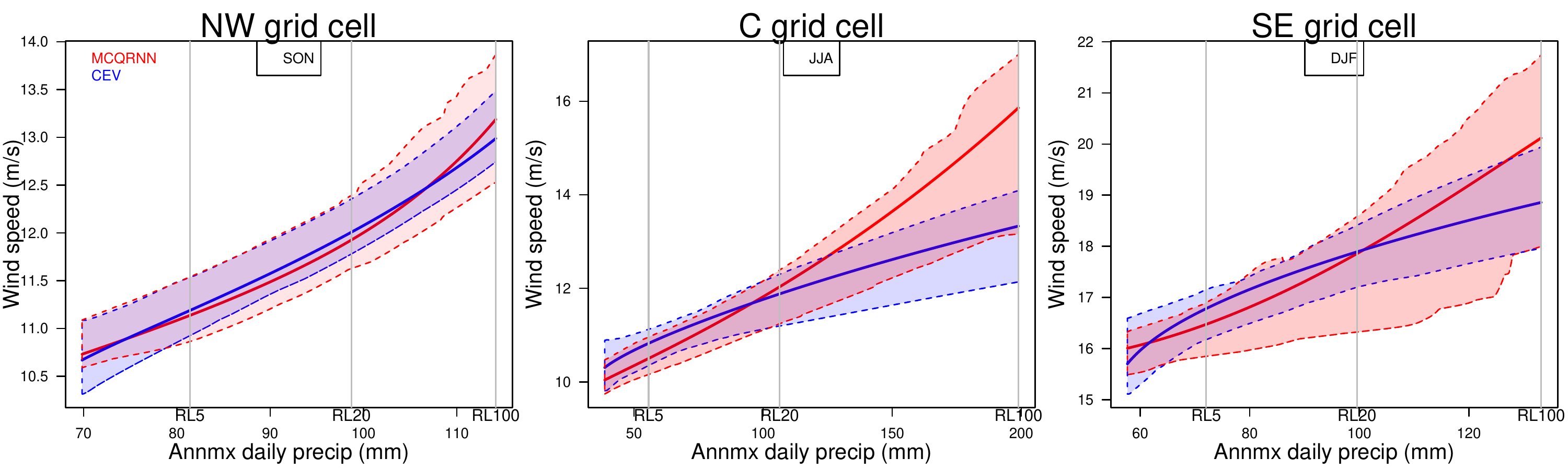}

    \caption{The estimated 0.9 conditional quantile curves (using the all the original ensemble members, solid blue (CEV) / red (MCQRNN) lines) and the 2.5/97.5 point-wise percentiles based on 100 ensemble bootstrapped samples (dashed lines)  at NW during SON (\textbf{left}), C during JJA (\textbf{Middle}), and SE during DJF (\textbf{Right}). The vertical dashed lines indicate estimates of the 5-, 20-, 100-year return levels of their seasonal maximum precipitation.}
    \label{fig:boot}
\end{figure}

 Having multiple ensemble members not only enables us to obtain more precise estimates of the distribution of conditional extremes but also allows a straightforward assessment for the estimation uncertainty by using the bootstrap \citep{efron1979}. Specifically, we can bootstrap the ensemble members to create a bootstrapped sample since to a good approximation each ensemble member is an independent realization of the underlying climate process. Fig.~\ref{fig:boot} shows the estimates of the conditional 0.9 quantile based on 100 bootstrapped samples for NW during SON, C during JJA, and SE during DJF. These results reconfirm the finding of the simulation study that the MCQRNN estimator becomes less stable (i.e., has larger estimation variation) for more extreme values (e.g., 100-year return level) of seasonal maximum precipitation. Because of the large dataset provided by the large ensemble it is possible to explore the deep upper tail and  demonstrate the uncertainty of MCQRNN estimates for these values.    

\begin{figure}[H]

    \centering
    \includegraphics[width=4.5in]{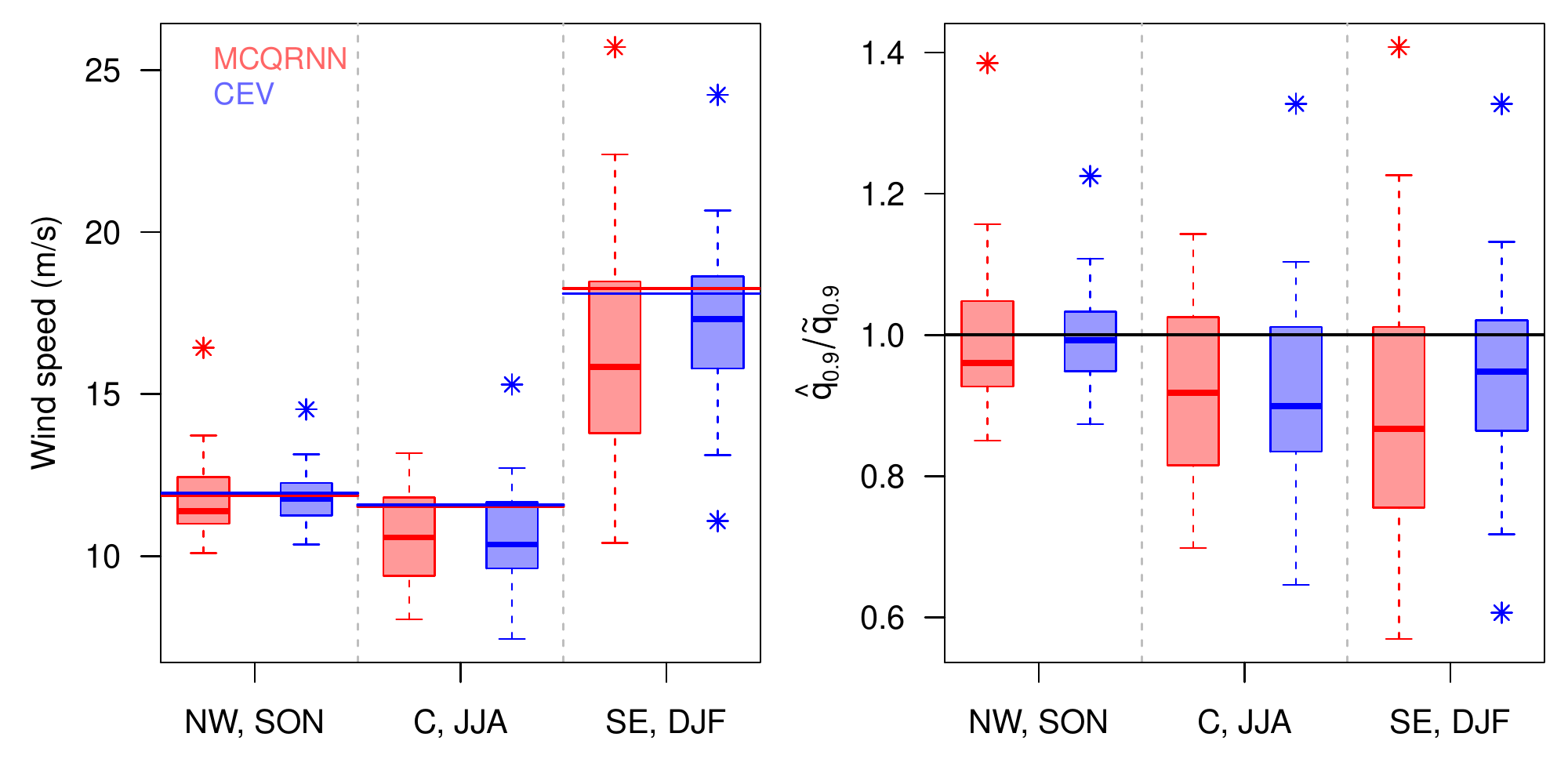}
    \caption{\textbf{Left}: Boxplots of conditional 0.9 wind speed quantile estimates at NW (SON), C (JJA), and SE (DJF) from all individual ensemble members using both MCQRNN and CEV. The horizontal lines are the corresponding 0.9 quantile estimates (red for MCQRNN and blue for CEV) estimated from the full ensemble. \textbf{Right}: Boxplots of conditional 0.9 quantile estimates divided by the MCQRNN estimates using all the ensembles combined.}
    
    \label{fig:pred}
\end{figure}

Another benefit of a large ensemble for studying climate extremes is that one can use estimates obtained using all the ensemble members as the ``truth'' to assess the precision and accuracy of estimates from smaller datasets more representative of the observational record. As an example, we consider the conditional 0.9 quantile of the distributions of daily wind speed at the NW (during SON), C(during JJA), SE (during DJF) grid points given that their concurrent seasonal maximum daily precipitation amount equals 96.76 mm (18-yr RL), 90.69 mm (13.8-yr RL), and 106.15 mm (27.6-yr RL) respectively. These values were chosen as they correspond to the largest values in which the MCQRNN and CEV estimates are still ``consistent'' (Fig.\ref{fig:boot}). We assess the estimation performance of MCQRNN and CEV by comparing the estimates for each individual ensemble with the associated conditional quantile estimates using all ensemble members (Fig.~\ref{fig:pred}, horizontal lines). As was found for most copulas considered in the simulation study, both MCQRNN and CEV underestimate the conditional 0.9 quantile. The magnitude of the underestimate tends to be larger for MCQRNN than CEV. Also, similar to what have found in our simulation study with sample size comparable to that of the data being considered, MCQRNN tends to be somewhat less stable: the MCQRNN standard errors are larger than those of CEV (see Table~\ref{tab:tab2}). Finally, these results demonstrate substantial spatial variability in the accuracy and precision of both methods.  At the NW grid box, characterized by relatively strong conditional dependence, both the bias and sampling range are relatively small compared to the ``true'' value obtained by using all the ensemble members. In contrast, at the SE grid box, for which the conditional dependence is relatively weak, both the bias and sampling range are quite large. The implication here is that one may prefer to use the CEV method with limited data (e.g., several decades of observations or model simulation) given the generally reasonable performance of the CEV model. The results here also demonstrate the value of using all of the data together in the analysis
at once, rather than evaluating each ensemble member individually and then using some measure of central tendency
across these results

\begin{table}[H]
\centering
{\small\begin{tabular}{l|c|c|c|c|c|c}
    & \multicolumn{2}{c}{NW (SON)}& \multicolumn{2}{c}{C (JJA)} &\multicolumn{2}{c}{SE (DJF)}\\
     \cline{2-7}
    (m/s) & MCQRNN&CEV&MCQRNN&CEV &MCQRNN&CEV\\
     \hline
    BIAS & -0.24  & -0.19 & -1.26 & -0.92 & -1.71& -1.15\\
    SE & 1.28&0.85& 1.48&1.44&2.37&3.51\\
    RMSE &1.30& 0.87& 1.94& 1.70& 3.90 & 2.91\\
     \hline
\end{tabular}}
\caption{The 0.9 wind speed quantile estimation bias (BIAS), standard error (SE), and root mean square error (RMSE) for MCQRNN and CEV. BIAS and RMSE values are calculated using the differences between the means of the single-realization estimates for both CEV and MCQRNN with the ``ground truth'' obtained by using MCQRNN. Standard error values are based on the range of the values estimated from individual ensemble members.}
\label{tab:tab2}

\end{table}

\section{Summary and Discussion} \label{sec6}

In this work we propose a conditional framework for estimating concurrent climate extremes in the bivariate setting. This approach allows for estimation of the entire distribution of one variable given that the other takes an extreme value, rather than just the joint probability of both variables being extreme. The estimation is conducted by decomposing the bivariate distribution into the marginal (tail) distribution of the conditioning variable and the conditional distribution of the concomitant values and modeling these separately. We employed univariate extreme value model for marginal modeling of the conditioning variable and we explore a non-parametric quantile regression method, the MCQRNN \citep{cannon2018}, and the conditional extreme value model \citep{heffernan2004,keef2013} for estimating the conditional distributions. We first illustrate this framework by conducting a Monte Carlo study and then applying it to estimate conditional high quantiles of daily wind speed given daily precipitation being seasonal maxima, using output from the CanRCM4 large ensemble. The results from both our simulation study and the specific application show that, despite both methods generally slightly underestimating the conditional upper quantiles, the estimates are reasonable (i.e., the biases are small compare to the magnitudes of the conditional upper quantiles) when the value of the conditioning variable is not too large. However, both this bias and the estimation variance generally amplify for the far tail of the conditioning variable, reflecting the fact that it is very difficult to make a reliable estimate with very limited data. The climate model large ensemble also allows for an assessment of estimation uncertainty using a straightforward ensemble-member bootstrap, and quantifying estimation bias and variance with data length comparable with a single model run as demonstrated in Sec.~\ref{sec5}.

Due to the lack of a theoretically-based asymptotic distributional form for the conditional distribution, it is still difficult to estimate the conditional high quantiles for the far tail of the conditioning variable even with our large ensemble. Due to its non-parametric nature, the MCQRNN method suffers from larger estimation variation and potentially larger estimation bias compare to that of the CEV when avaiable data are limited. The CEV model, on the other hand, can gain estimation efficiency if the model assumptions are reasonable, but may perform poorly when these assumptions are violated. While estimates obtained by using CEV methods tend to be more stable, they may suffer from model misspecification as illustrated in Sec.~\ref{sec4}. Having two estimates allows one to identify potential issues for a follow-up investigation.

There are natural directions of future work. First, rather than using block maxima, it would be natural to use threshold exceedances and their concomitant values for estimating the marginal tail distributions and the conditional distribution. Second, the variables of interest often exhibit coherent spatial structures. Therefore, it is sensible to \textit{assume} there exists a smooth spatial pattern in terms of the (extremal) dependence structure between variables. Some form of spatial smoothing could potentially improve the estimation to the extent that sampling error in space is a substitute for sampling error in time. Third, it would be interesting to study of the conditional upper tail of daily precipitation conditioning on extreme wind speed, to illustrate the symmetry in terms of the dependence structures (or the lack of such symmetry). A technical complication associated with such an analysis is that the concomitant distribution of precipitation conditioned on extreme wind speed is a mixture of a continuous distribution with a point mass at zero.

Fourth, large ensembles allow for the investigation of the use of non-stationary marginal and dependence structures for extremes, which is crucial for climate change studies.
An important task here is to 
first understand and then to appropriately parameterize the nonstationary dependence structure in a way that is both parsimonious and yet flexible enough to approximate the true dependence structure. Finally, the proposed framework can be applied to the whole spatial domain to characterize the spatial nature of concurrent extremes.     

\section*{Acknowledgements} \label{acknowledgements}
This work was conducted as part of the Canadian Statistical Sciences Institute (CANSSI, \url{http://www.canssi.ca/}) Postdoctoral Fellowships program. We acknowledge the Canadian Center
for Climate Modeling and Analysis of
Environment and Climate Change
Canada for executing and making
available the CanRCM4 large ensemble
simulations. WH acknowledges the support support of the NSF Grant \# 1638521 to the Statistical and Applied Mathematical Sciences Institute (SAMSI). AM acknowledges the support of the Natural Sciences and Engineering Research Council of Canada (NSERC) [funding reference number RGPIN-2019-04986]. The authors would also like to thank Dr.\ Alex Cannon and two anonymous reviewers for their valuable input.


\section*{Appendices}

\begin{appendices}
\renewcommand{\thesection}{\Alph{section}}
\renewcommand{\thefigure}{A.\arabic{figure}}

\section{\Large{Seasonal Maximum Precipitation and Their Concomitant Wind Speed}}
\label{App:hists}

\setcounter{figure}{0}  
\begin{figure}[H]
    \centering
    \includegraphics[width=3in]{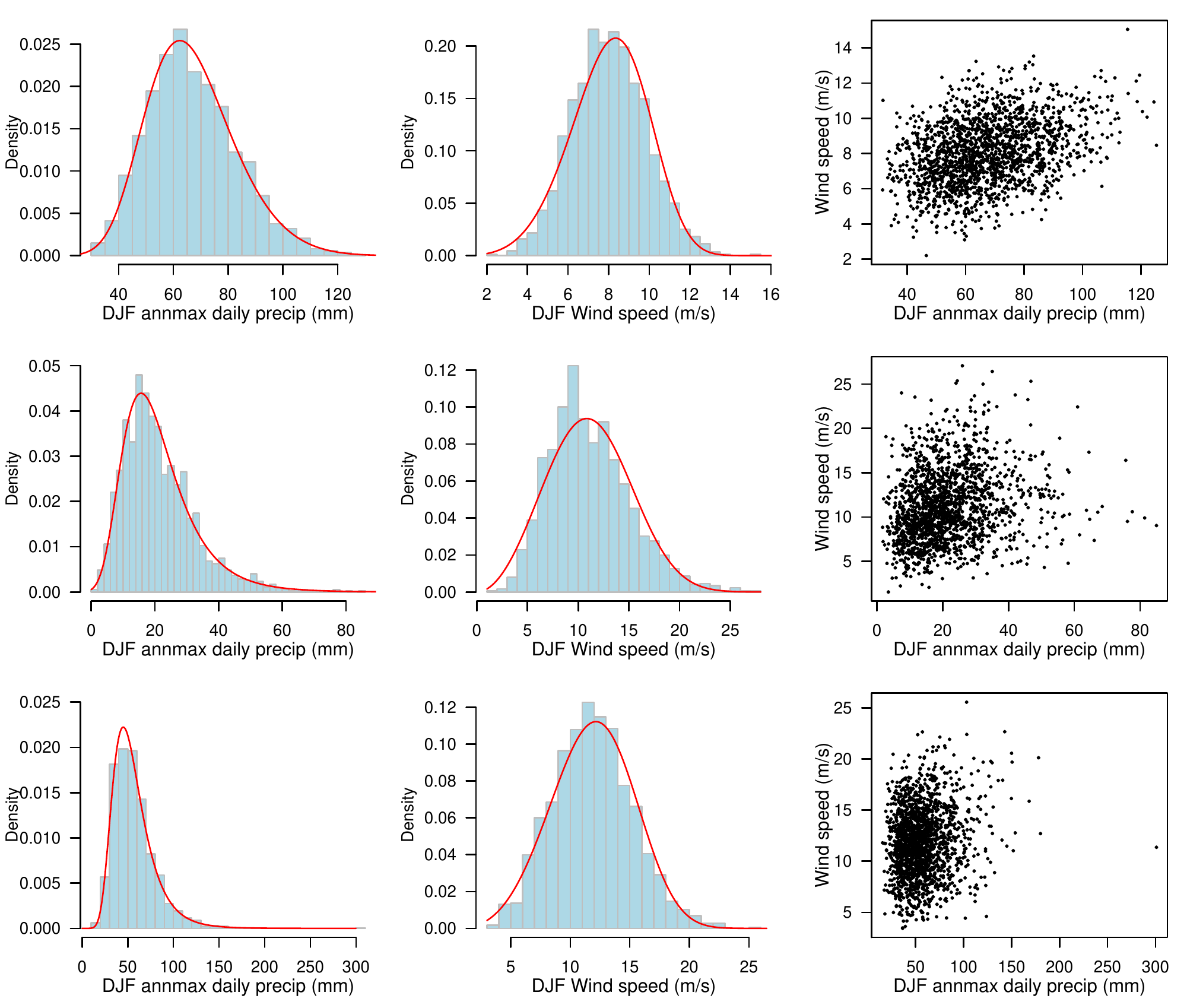} \hspace{0.2in}
    \includegraphics[width=3in]{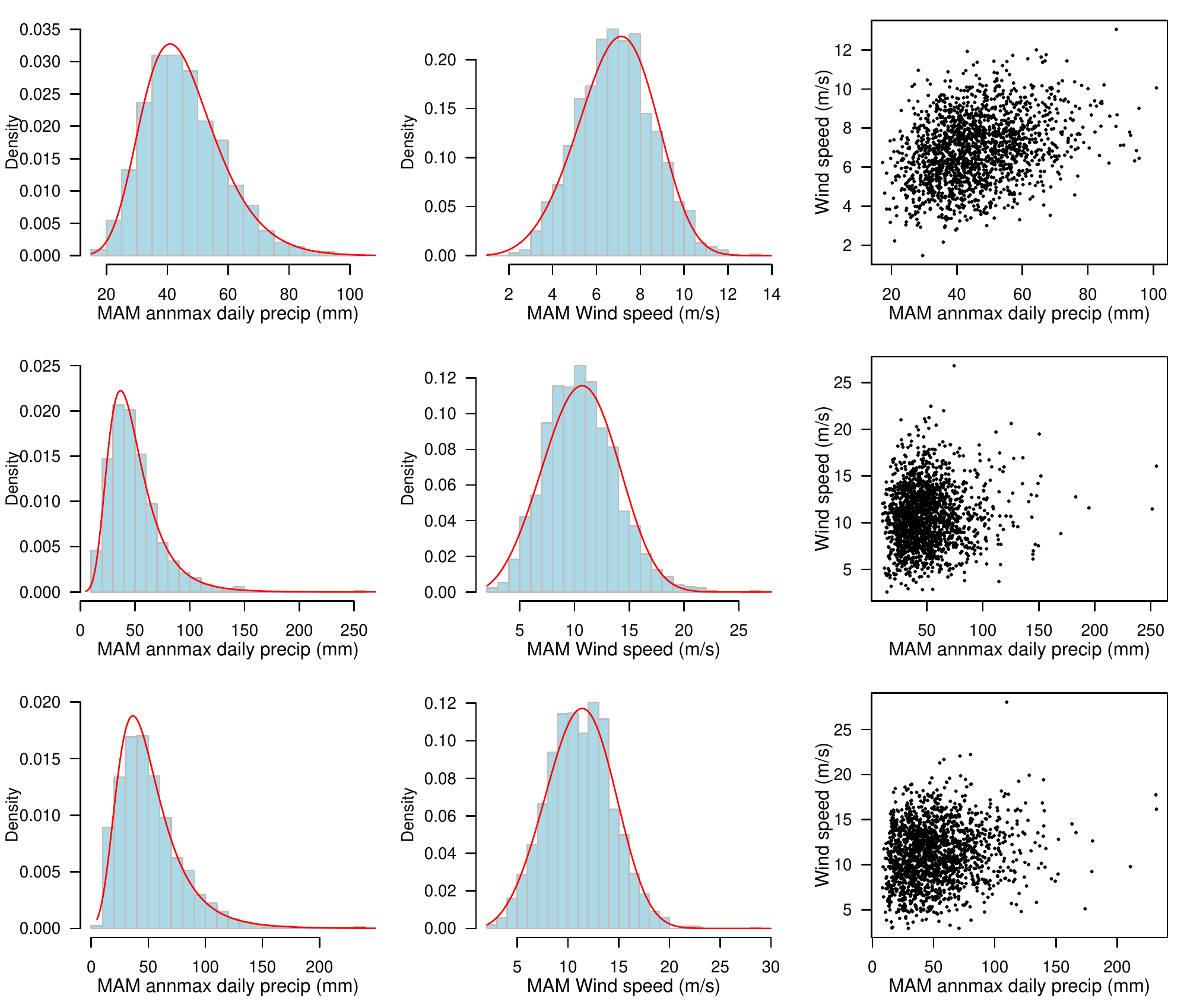}\\
    \vspace{0.2in}
    \includegraphics[width=3in]{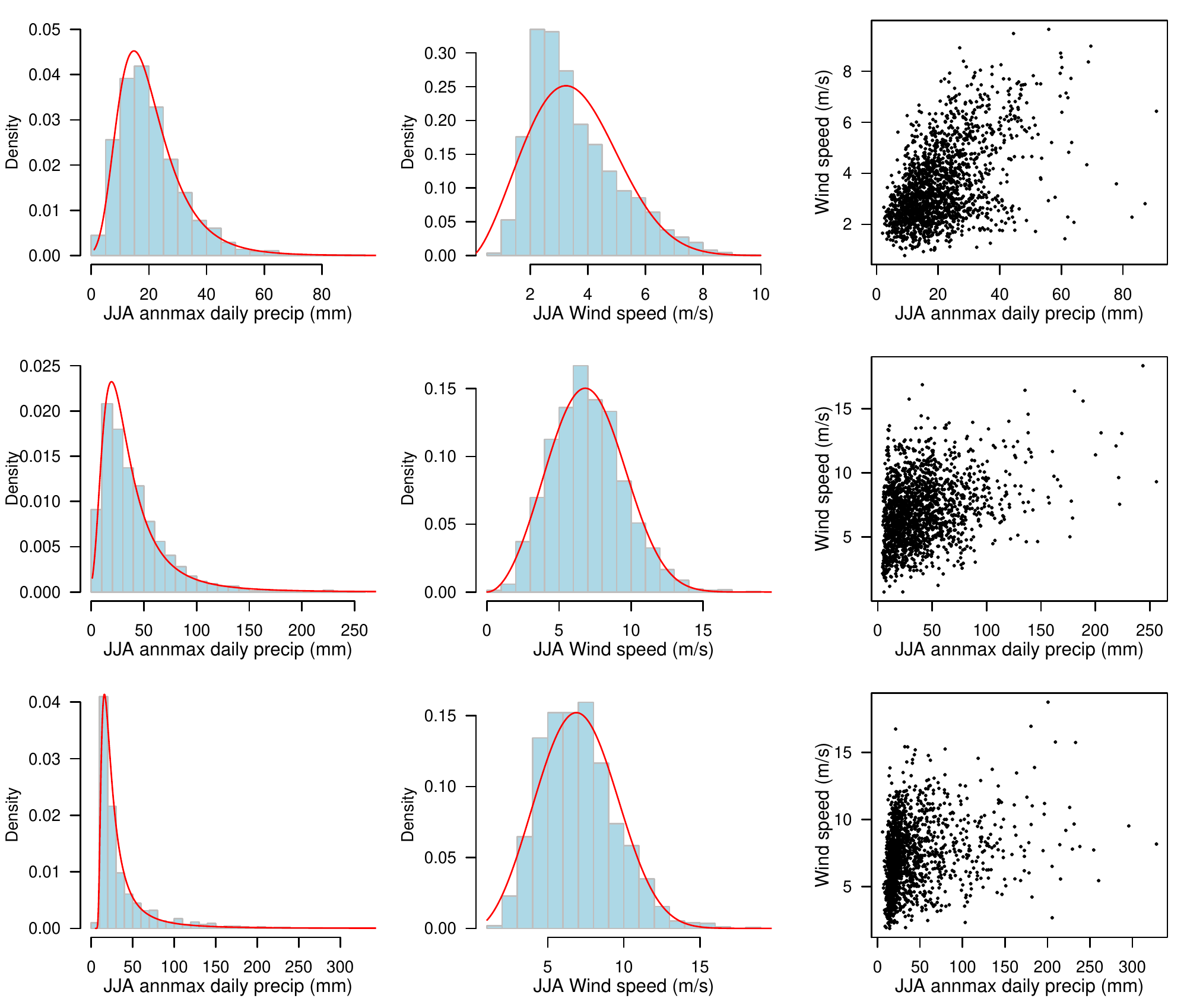}
    \hspace{0.2in}
    \includegraphics[width=3in]{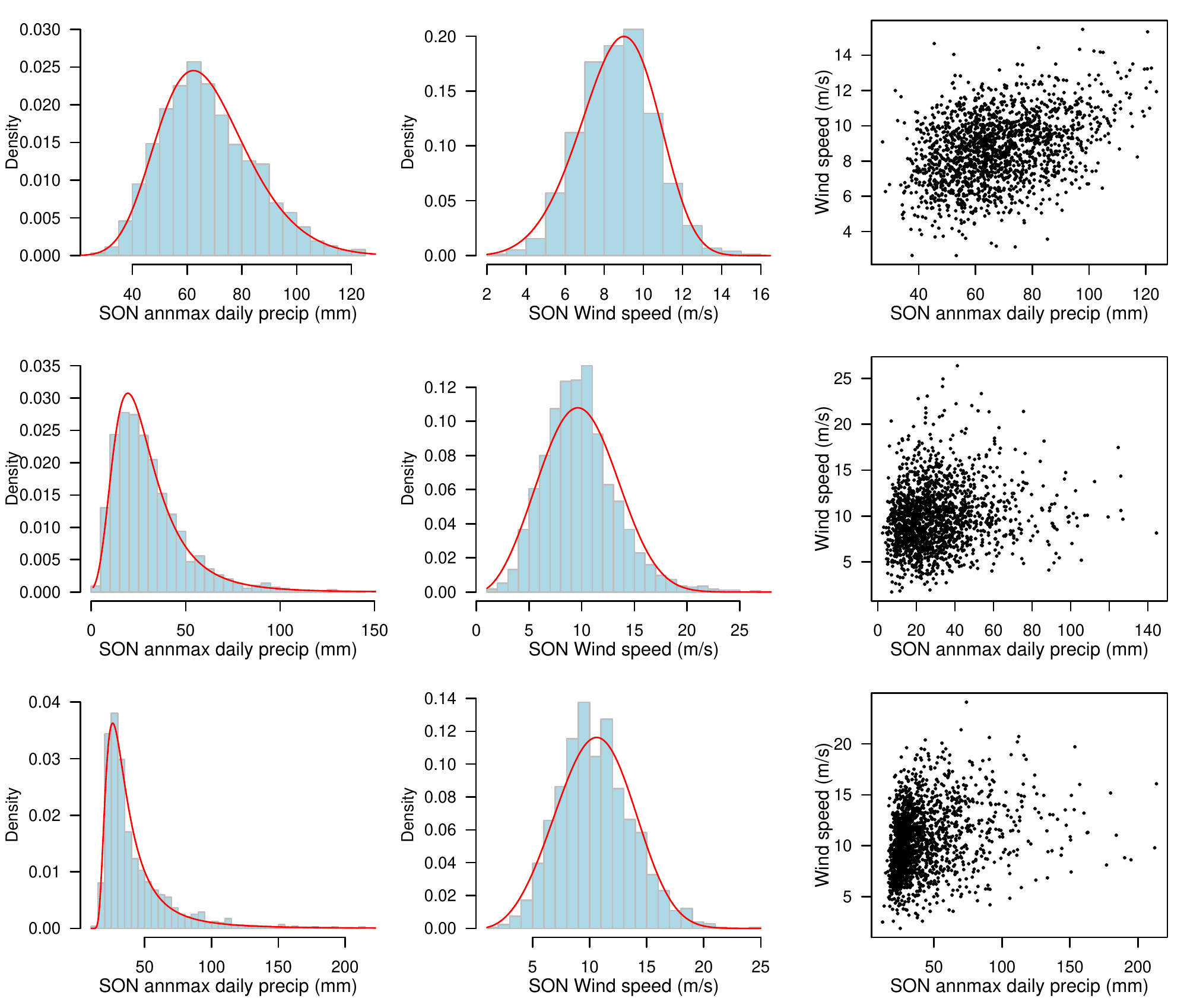}
    \caption{\textbf{Top row}: Histograms of seasonal maximum daily precipitation (\textbf{Left}), the concomitant daily average wind speed (\textbf{Middle}), and the scatterplot of both together (\textbf{Right}) for a Pacific Northwest (NW) grid cell. \textbf{Middle row}: As in the top row but a Continental interior (C) grid cell. \textbf{Bottom row}: As in the top row but for a Southeastern (SE) grid cell. The red curves are the fitted densities (GEV for annual maximum precipitation and Weibull for concomitant wind speed).}
    \label{fig:histDJF}
\end{figure}

\renewcommand{\thefigure}{B.\arabic{figure}}

\setcounter{figure}{0}  

\section{Annual Maximum Daily Precipitation and Their Concurrent Daily Average Wind Speed 1950-2100}
\label{App:nonstat}

\begin{figure}[H]
    \centering
    \includegraphics[width=5in]{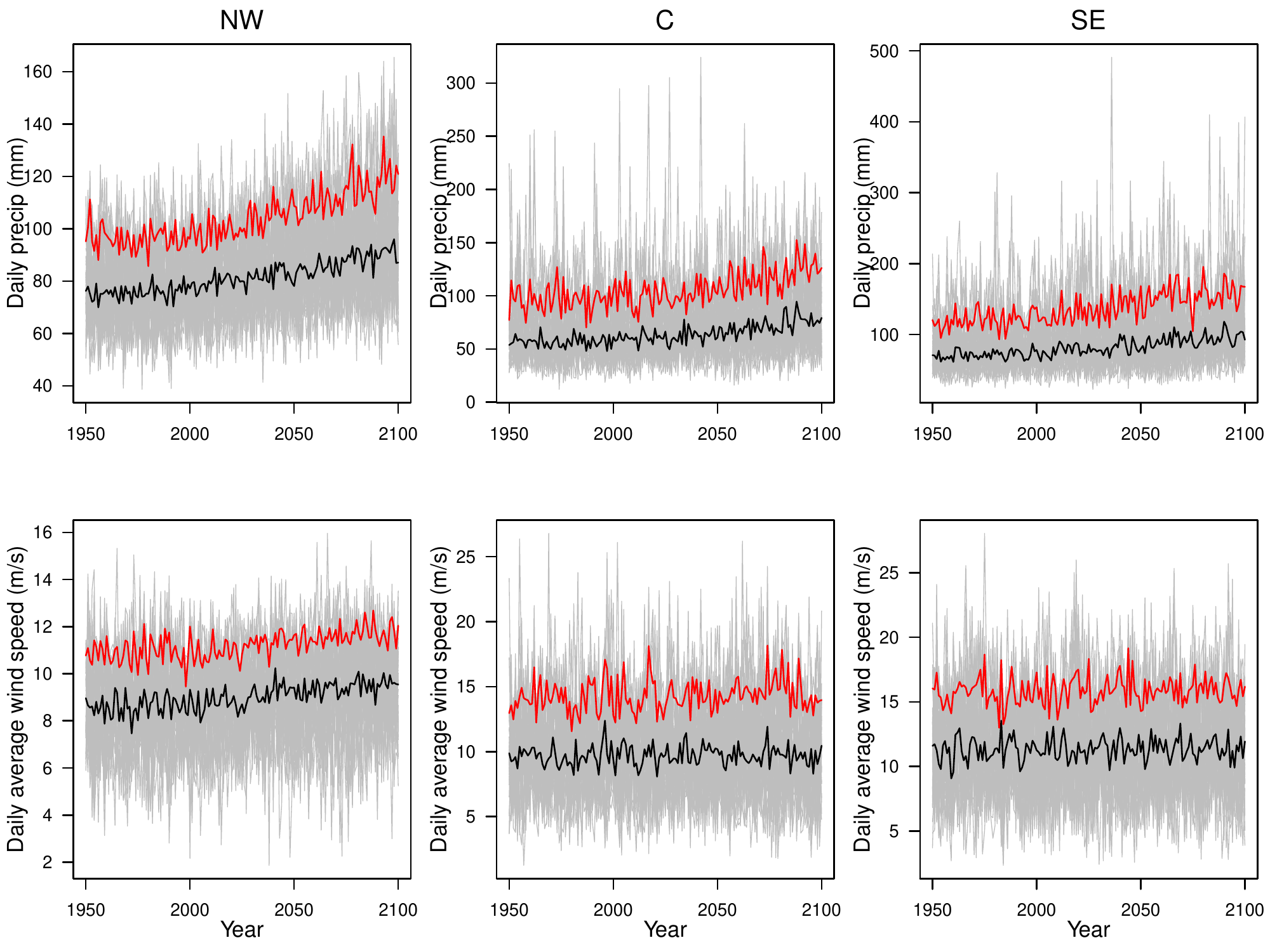}
    \caption{\textbf{Top:} The time series of annual maximum daily precipitation at NW, C, and SE grid cells. \textbf{Bottom:} The concurrent daily average wind speed values. The black lines are the ensemble median (for each year) and the red lines are the ensemble 0.9 quantile.}
    \label{fig:nonstatTs}
\end{figure}

\renewcommand{\thefigure}{C.\arabic{figure}}

\setcounter{figure}{0}  

\section{Model Fitting Procedures}\label{App:fit}

In this work all statistical inference was performed using frequentist approaches. Specifically, the estimation of $\alpha$ and $\beta$ in CEV is based on a Gaussian working assumption on the distribution of $Z = \frac{Y_{2} - \alpha Y_{1}}{Y_{1}^{\beta}}$ given $Y_{1}>u$. Our implementation is based on the function \texttt{mex} in \texttt{texmex} R package \citep{texmex}. Estimation uncertainty can be calculated using a bootstrap procedure as suggested in \cite{heffernan2004}. In this work we exploit the large ensemble and apply an \textit{ensemble-bootstrap} \citep[][i.e., bootstrap across ensemble members]{haugen2018}  to
estimate the uncertainty of the estimate obtained using all the ensemble members. For the \texttt{MCQRNN} we make use the \texttt{mcqrnn.fit} function in the R package \texttt{qrnn}; the details can be found at \cite{cannon2018}.
\label{App:Fit}

\renewcommand{\thefigure}{D.\arabic{figure}}

\setcounter{figure}{0}

\section{Estimation Uncertainty using All the Ensemble Members}
\label{App:UQALL}
\begin{figure}[H]
    \centering
    \includegraphics[width=6in]{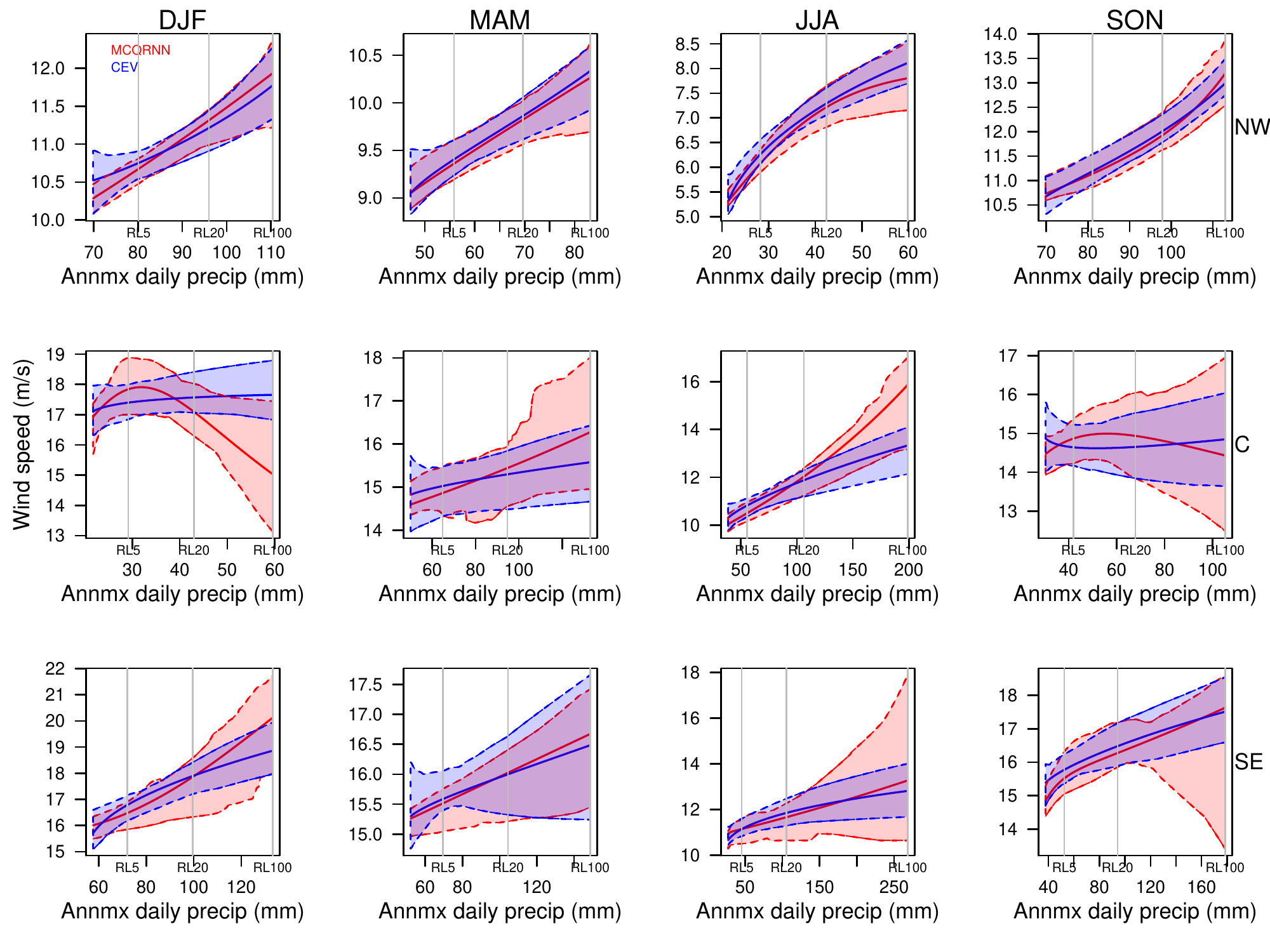}
    \caption{As in Fig.~\ref{fig:boot} but for all season and grid cell combinations.}
    \label{fig:UQ_fbplotAll}
\end{figure}

\renewcommand{\thefigure}{E.\arabic{figure}}

\setcounter{figure}{0} 

\section{Explaining Negative Biases of Upper Conditional Quantiles}
\label{App:bias}

\begin{figure}[H]
    \centering
\includegraphics[width=3.5in]{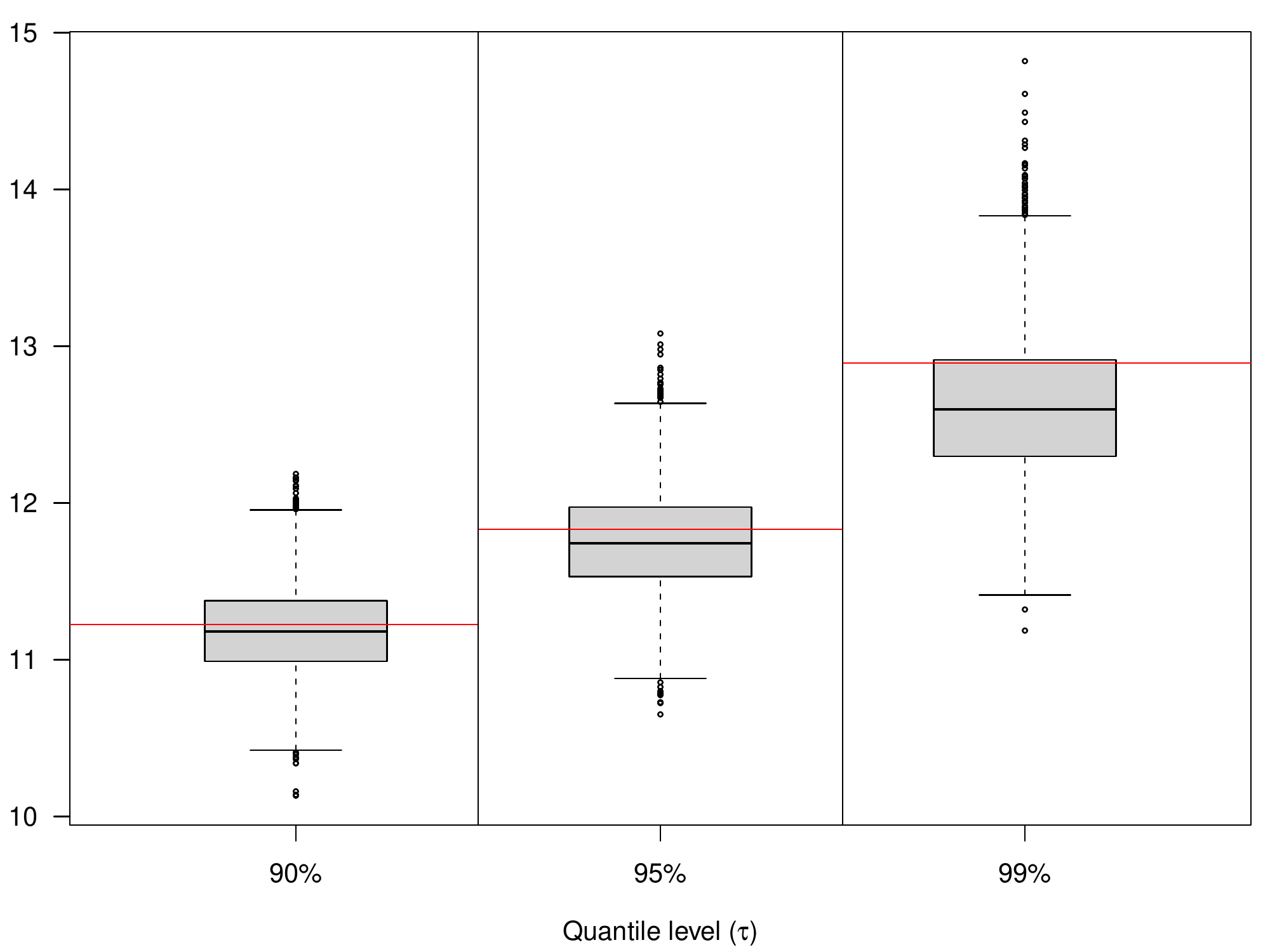}
    \caption{Boxplots of empirical estimates of .9, .95, and .99 quantiles for all 5,000 Monte Carlo simulations with sample size $n=100$. The red horizontal lines are the true quantile values.}
    \label{appfig4}
\end{figure}

The negative bias issue arises due to the method used to estimate high empirical quantiles (e.g., $\tau=$ 0.9) in the implementations of both MCQRNN and CEV. MCQRNN and CEV are affected differently because MCQRNN estimates high quantiles ``locally'' while CEV estimates quantiles ``globally'' after standardizing with the location and scale parameters (i.e., $\alpha$ and $\beta$) of the dependence structure. Below we provide a simulation study to demonstrate the negative bias issue that arises with the quantile estimating method that is embedded in the \texttt{R} packages that provide the implementations of the CEV and MCQRNN we employ, which is the default method in the \texttt{quantile} command in the \texttt{stats} \texttt{R} base package \citep[see][ for more details]{hyndman1996}. In the first setting, we assume that the true wind speed distribution conditioned on precipitation amount being annual maximum follows a Weibull distribution with shape parameter $5$ and scale parameter $9.5$ (these values are MLE parameter estimates at NW (SON)). We simulate 5,000 replicates, each of which consists of 100 data points (the sample size $n$), compute empirical estimates of the 0.9, 0.95, and 0.99 quantiles and compare these empirical estimates with the true quantile values. The result in Fig.~\ref{appfig4} shows that we have negative biases for all the cases and that the bias increases with quantile level (for a fixed sample size).

\begin{figure}[H]
    \centering
\includegraphics[width=3.5in]{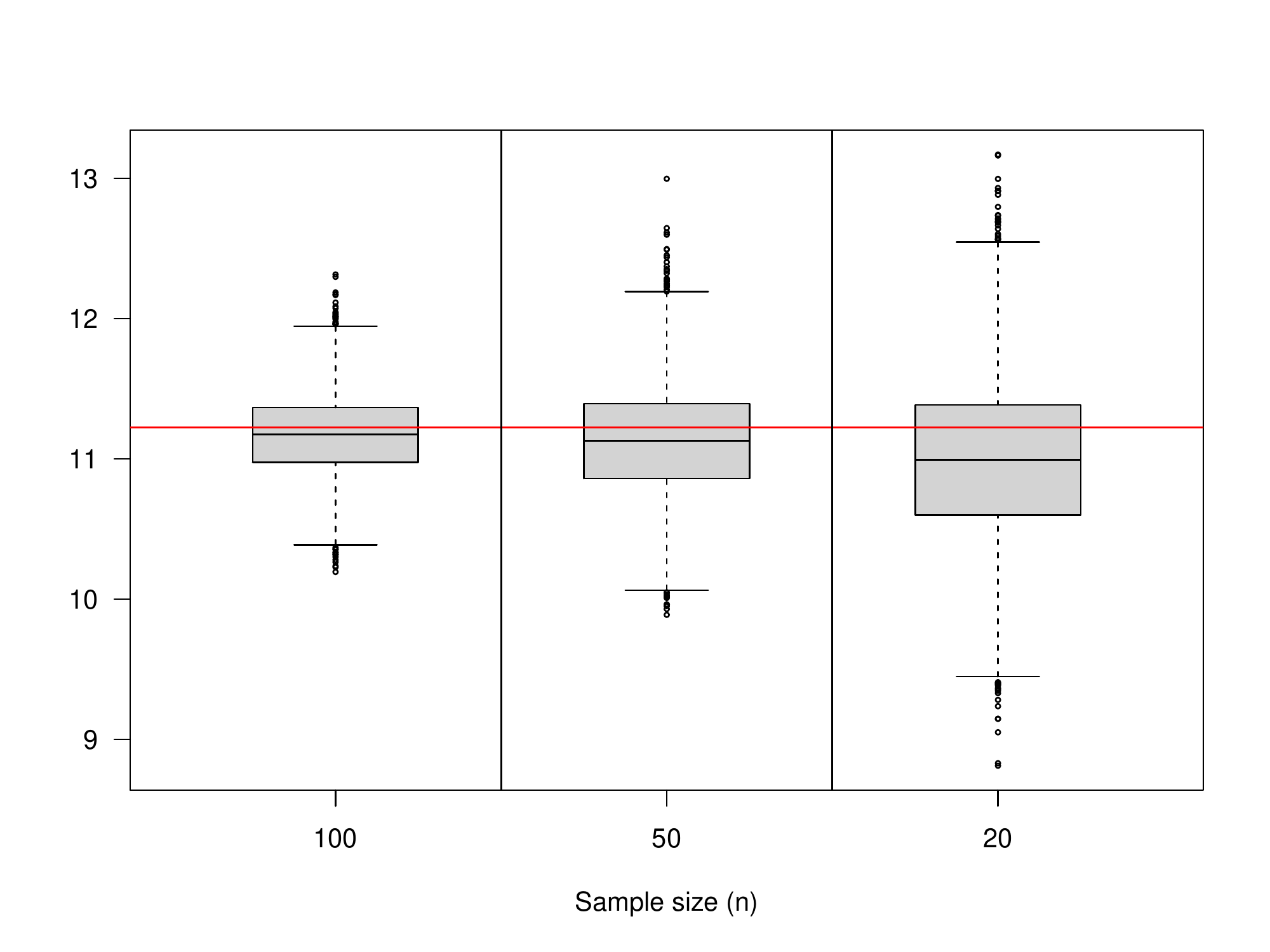}
    \caption{Boxplots of empirical estimates of .9 quantile for all 5,000 Monte Carlo simulation with sample sizes $n=100$ (Left), $n=50$ (Middle), and $n=20$ (Right). The red horizontal line is the true quantile value.}
    \label{appfig5}
\end{figure}

In the second setting, we focus on estimating the 0.9 quantile but we change the sample size $(n = 100, 50, 20)$ for each of those 5,000 replications to reflect the fact that we would like to non-parametrically estimate the nonlinear quantile curves at the same time, which implies that the ``effective'' sample size for estimating quantiles locally is smaller. The result in Fig.~\ref{appfig5} shows that, for a given quantile level (0.9 here), the negative bias increases with decreasing sample size. For the estimation of the 0.9 conditional quantile of wind speed given precipitation amount being annual maximum, the use of parametric quantile curves in CEV (i.e., the location and scale parametric form after marginal transformation) results in larger effective sample sizes (compared with MCQRNN) for estimating the high quantiles. It is also worth noting that CEV quantile estimation is not entirely based on empirical quantiles as its marginal distribution has been modeled as a mixture of an empirical estimate and a generalized Pareto tail.

The negative biases shown here are partly  due the default implementation of quantile estimation in \texttt{R}. Specifically, all sample quantiles are defined as weighted averages of consecutive order statistics,  with more weight placed on the lower order statistic, which causes the issue of underestimation. 

\end{appendices}

\singlespacing
\bibliography{CompoundEx.bib}

\end{document}